\documentclass{appolb}
\usepackage{graphicx}
\usepackage{amsmath,amssymb}
\usepackage{aas_macros}
\usepackage{booktabs}

\usepackage[style=numeric-comp,sorting=none,maxnames=4]{biblatex}
\usepackage{enumitem} 

\graphicspath{{./figures/}}

\newcommand{\Tint}[1]{{\hbox{$\sum$}\!\!\!\!\!\!\!\int\,}_{\!\!\!\!\raise-0.9ex\hbox{$\scriptstyle{#1}$}}}

\begin{document}
\title{Particle-theory input for neutron-star physics%
\thanks{Presented at the 63rd Cracow School of Theoretical Physics, 17-23 September 2023.}%
}
\preprint{\hspace*{0pt}\hfill HIP-2024-14/TH}
\author{
{Aleksi Vuorinen
\address{Department of Physics and Helsinki Institute of Physics, P.O.~Box 64, FI-00014, University of Helsinki, Finland}
}
}
\maketitle
\begin{abstract}
Understanding the properties and physical phase of the dense strongly interacting matter present in the cores of neutron stars or created in their binary mergers remains one of the most prominent open problems in nuclear astrophysics. While most microscopic analyses have historically relied on solvable phenomenological models of nuclear and quark matter, in recent years a model-independent approach utilizing only controlled \textit{ab-initio} calculations and astrophysical observations has emerged as a viable alternative. 

In these lecture notes, I review recent progress in first-principles weak-coupling calculations within high-density quark matter, shedding light on its thermodynamic and transport properties. I cover the most important technical tools used in such calculations, introduce selected highlight results, and explain how this information can be used in phenomenological studies of neutron-star physics. The notes do not offer a self-consistent treatment of the topics covered, but rather aim at filling gaps in existing textbooks on thermal field theory and at connecting the dots in a story developed in several recent research articles, to which the interested reader is directed for further technical details. 
\end{abstract}
  
\section{Introduction}

The qualitative idea that a competition between gravity and the degeneracy pressure of matter composed of nucleons could give rise to compact astrophysical objects dates back to more than 90 years \cite{Landau:1932uwv}, and the first direct observation of rapidly rotating pulsars will soon reach the ripe age of 60 \cite{Hewish:1968bj}. Yet, it is only during the past 10 years that neutron stars (NSs) have truly become a functioning laboratory of dense nuclear and quark matter (NM and QM), owing largely to recent dramatic advances in observational astrophysics. Milestone results from the 2010s and 2020s include accurate mass measurements of several individual high-mass NSs \cite{Antoniadis:2013pzd,Cromartie:2019kug,Fonseca:2021wxt}, often taking advantage of General Relativistic effects such as the Shapiro Delay \cite{Shapiro:1964uw}; increasingly precise radius measurements utilizing X-ray emission from NSs \cite{Shawn:2018,Nattila:2017wtj,Miller:2019cac,Miller:2021qha,Riley:2019yda,Riley:2021pdl}; and most famously the first-ever detection of a gravitational-wave (GW) signal from a binary NS merger by the LIGO and Virgo collaborations in 2017 \cite{LIGOScientific:2017vwq,LIGOScientific:2018cki,LIGOScientific:2018hze}. In addition to the first GW signal GW170817, the same merger event gave rise to an associated electromagnetic signal across a wide spectrum, recorded by altogether 70 different observatories \cite{LIGOScientific:2017ync,LIGOScientific:2017zic}. This marked the dawn of an era of multimessenger astronomy, the future of which looks bright, with several new GW observatories capable of recording a postmerger signal, including the Einstein Telescope \cite{Maggiore:2019uih} and the Cosmic Explorer \cite{Reitze:2019iox}, being planned at the moment.

In order to draw robust microphysics lessons from the increasing amount of high-quality observational data on NS properties, it is imperative to simultaneously develop our theoretical understanding of matter at supernuclear densities. Here, the dominant physical effects are described by the theory of the strong nuclear interaction, Quantum Chromodynamics (QCD), with subleading but important roles played by the electromagnetic and weak sectors of the Standard Model. Due to the Sign Problem of lattice QCD at nonzero baryon chemical potentials $\mu_B$ \cite{deForcrand:2009zkb}, a nonperturbative first-principles approach available at all relevant densities is unfortunately unavailable, leaving us with effective-theory frameworks and weak-coupling expansions to work with. Indeed, the \textit{ab-initio} tools available in the cold and dense part of the QCD phase diagram include Chiral Effective Theory (CET), valid for moderate-density NM up to somewhat above the nuclear saturation density $n_s=0.16/$fm$^3$ (see, e.g., \cite{Machleidt:2011zz}); perturbative QCD (pQCD), available beyond some tens of saturation densities \cite{Laine:2016hma,Ghiglieri:2020dpq}; and with some reservations methods such as Functional Renormalization Group \cite{Pawlowski:2005xe} the AdS/CFT conjecture that allows access to the strongly coupled regime of many QCD-like theories albeit not QCD itself \cite{Casalderrey-Solana:2011dxg,Hoyos:2021uff}. 

In the lecture notes at hand, our focus will be in the physics of ultradense QM. More specifically, we aim to provide an introduction to the methods used in recent perturbative determinations of the thermodynamic and transport properties of dense unpaired QM and to briefly review the application of these results to phenomenological studies of NS matter. While we will at times perform brief example calculations, these notes are not meant to serve as a stand-alone textbook. On the contrary, we will rely on existing treatments of the basics of thermal field theory and supplement them with brief introductions to computational methods specific to finite density, pointing out the most useful references where they have been developed or applied at the state-of-the-order level. 

The remaining seven sections of the notes cover the following topics: 
\begin{enumerate}
\setcounter{enumi}{1}
    \item Basics of perturbative thermal field theory at finite density
    \item Soft effective theories at finite temperature and density
    \item The Hard Thermal Loop effective theory
    \item High-order perturbative calculations at finite density
    \item Transport properties of dense QM
    \item Model-independent studies of the NS-matter equation of state
    \item Outlook towards future developments
\end{enumerate}
In sections 2-6, we will pay particular attention to thermal-field-theory techniques specific to finite density that have received limited attention in existing literature, while in sec.~7, we will focus on the role of QM thermodynamics in recent model-independent determinations of the NS-matter equation of state (EoS).

In the perturbative calculations performed in these lecture notes, we shall mostly employ the imaginary-time formalism of thermal field theory, implying that our four-dimensional metric is Euclidean $\eta_{\mu\nu}=\delta_{\mu\nu}$. For all other conventions, we use the definitions specified in the preface of \cite{Laine:2016hma}.

\section{Basics of perturbative thermal field theory in dense QCD}

Perturbation theory is without doubt the most popular and versatile computational tool in quantum field theory (QFT), with applications ranging from collider problems in vacuum (see, e.g.~\cite{Collins:2011zzd}) to thermal-field-theory challenges in hot and/or dense environments such as the early Universe or the fireball created in a heavy-ion collision \cite{Ghiglieri:2020dpq}. Within the latter context, the weak-coupling approach can be used to determinate not only Euclidean "bulk thermodynamic" quantities but also transport coefficients \cite{Ghiglieri:2013gia,Ghiglieri:2018dib} and even time-dependent quantities related to, e.g., thermalization dynamics \cite{Kurkela:2011ub}. 

In the context of NS physics, of primary interest in these lecture notes, the phenomenologically most important quantity is the EoS of dense beta-equilibrated QCD matter, i.e.~the relationship between its pressure and energy density. This quantity can be shown to be in a rough one-to-one correspondence with the so-called NS mass-radius relation (see sec.~7 below), implying that both NS observations and microphysical calculations can be used in a model-independent inference of the properties of NS matter. Such calculations form the topic of sec.~7 below and serve as the primary motivation for the high-order perturbative determinations of QM thermodynamics that we will be studying in this and the following three sections.

In the realm of high temperatures, perturbative thermal field theory is a well-developed and mature field, with multiple textbooks devoted to the subject. For the imaginary-time formalism we employ here, two widely used sources include \cite{Kapusta:2006pm,Laine:2016hma}, while for the real-time alternative we can recommend \cite{Bellac:2011kqa,Ghiglieri:2020dpq}. Given that these references touch upon cold and dense fermionic matter only very briefly (two exceptions being sec.~7 of \cite{Laine:2016hma} and sec.~6.4 of \cite{Ghiglieri:2020dpq}), we will here start from a brief review of the basic equilibrium thermodynamic properties of such systems at low perturbative orders. 

Any weak-coupling calculation in cold and dense QM proceeds as an expansion around a system of noninteracting quarks at sizable chemical potentials and small or vanishing temperature. The physical properties of such a system resemble those of the conduction electrons in a metal --- a canonical example system in undergraduate statistical mechanics --- and can be qualitatively understood based on the small-$T$ behavior of the Fermi-Dirac distribution function
\begin{eqnarray}
n_{F}(\epsilon)&=& \frac{1}{e^{(\epsilon-\mu)/T}+1} \;\underset{T\to 0}{\longrightarrow} \; \theta(\mu-\epsilon)\, , 
\end{eqnarray}
with $\epsilon=\epsilon(p)$ denoting the free one-particle dispersion relation. The step-function form of this quantity implies that in the strict $T=0$ limit, all quantum-mechanical states are filled inside a momentum-space Fermi sphere with radius determined by the relation $\epsilon(p)=\mu$ and unoccupied outside it. This reflects the effect of the Pauli exclusion principle and is in stark contrast with the low-temperature behavior of bosonic systems, characterized by a condensation of particles to the lowest quantum state available \cite{Anderson:1995gf}.

Interaction corrections to the thermodynamic properties of the free system can be obtained by expanding the path-integral representation of the grand canonical partition function in powers of the coupling constant. For a relativistic gauge field theory coupled to Dirac fermions with separately conserved number densities (thus allowing the introduction of the corresponding chemical potentials),\footnote{In the case of QCD, this implies neglecting the effects of flavor-changing weak interactions, which is a good approximation in heavy-ion collisions, but not inside neutron stars. We will return to this issue when discussing beta equilibrium below.} this quantity takes the form \cite{Laine:2016hma}
\begin{eqnarray}
 \mathcal{Z}_\text{QCD} & = &  
 \int_\text{periodic} \!\! \mathcal{D} A_0^a \, \mathcal{D} A_k^a
 \int_\text{periodic} \!\! \mathcal{D} \bar c^{\,a} \, \mathcal{D} c^a 
 \int_\text{anti-periodic} \!\! \mathcal{D} \bar\psi_f \, \mathcal{D} \psi_f
\nonumber  \\ &\times&   
 \exp\bigg\{
  - \int_0^\beta \! {\text d}\tau \! \int_\mathbf{x} \,  
  \bigg[
   \frac{1}{4} F^a_{\mu\nu}F^a_{\mu\nu}
   + \frac{1}{2\xi} G^a G^a
   + \bar c^{\,a} \big( \frac{\delta G^a}{\delta \theta^b}\big) c^b  \nonumber \\
  &&\;\;\;\;\;\;\;\;\;\;\;\;\;\;\;\;\;\;\;\;\;\;\;\;\;\;\;+\, \bar \psi_f ( \gamma^{ }_\mu D^{ }_\mu + m_f -\mu_f \gamma_0) \psi_f   
  \bigg] 
 \bigg\} 
 \, . \label{path}
\end{eqnarray}
Here, we denote gluon fields in the adjoint representation of the SU(3) gauge group by $A_\mu^a$, $a=1,2,...,8$ being the corresponding color index; adjoint-representation ghosts by $c^a$: fundamental-representa\-tion quarks of flavor $f$ by $\psi_f$; the covariant gauge parameter by $\xi$; and the function defining the covariant gauge by $G^a \equiv -\partial_\mu A_\mu^a$. Finally, the field strength tensor and the fundamental-representation covariant derivative appearing above read
\begin{eqnarray}
F_{\mu\nu}^a&=&\partial_\mu A_\nu^a -\partial_\nu A_\mu^a + g f^{abc} A_\mu^b A_\nu^c\, , \quad D_\mu \;=\; \partial_\mu-igA_\mu^a T^a\, ,
\end{eqnarray}
where $f^{abc}$ and $T^a$ stand for the structure constants and Hermitean generators of the gauge group,  Boundary conditions of the fields in the imaginary time direction $\tau$, running from 0 to $\beta\equiv 1/T$, are indicated in the functional integral above, and the Euclidean $\gamma$ matrices and other necessary quantities are properly defined and listed in \cite{Laine:2016hma}.

Just as at high temperatures, the weak-coupling expansion of the pressure, or the logarithm of the partition function, of a cold and dense system is organized in terms of a loop expansion in connected vacuum or bubble Feynman diagrams. The first two orders of the expansion proceed without complications, with the subtraction of $\mu$-independent vacuum terms and the renormalization of quark masses and the gauge coupling $g$ sucecssfully removing all $1/\epsilon$ ultraviolet (UV) divergences encountered in dimensional regularization. Analogously to the high-$T$ case, uncancelled infrared (IR) divergences, however, appear at the three-loop order, which necessitates the use of either explicit diagrammatic resummations or low-energy effective field theories (EFTs). Given that this will be the main topic of the next two sections of these notes, we will leave this issue aside here and simply point out some of the most important technical differences between loop calculations performed in the high- and low-$T$ realms, relying on the reader's familiarity with the former context (based on textbooks such as \cite{Laine:2016hma}):
\begin{itemize}
\item While the use of the Matsubara formalism is required at all nonzero values of temperature in the imaginary-time formalism, both bosonic and fermionic sum-integrals become continuous integrals in $D=4-2\epsilon$ dimensions in the strict $T\to 0$ limit. This implies that we may write
\begin{eqnarray}
\Tint{P} &\rightarrow& 
\int_{-\infty}^\infty\frac{{\text d}p_0}{2\pi} \int  \frac{{\text d}^d\mathbf{p}}{(2\pi)^d} \; \equiv\;  \int_{-\infty}^\infty\frac{{\text d}p_0}{2\pi}
 \int_{\textbf{p}} \; \equiv\; 
 \int_{P}
  \,  \quad \mbox{(bosons)}\,,  \nonumber\\
  \Tint{\{P\}} &\rightarrow& 
  \int_{-\infty+i\mu}^{\infty+i\mu}\frac{{\text d}p_0}{2\pi}
 \int  \frac{{\text d}^d\mathbf{p}}{(2\pi)^d} \; \equiv \; 
 \int_{\widetilde{P}}
  \,  \quad \mbox{(fermions)} \,, \nonumber
\end{eqnarray}
where the tilde in the latter integration measure reminds us of the fermionic nature of the corresponding momentum. In practical calculations, it is often advantageous to start the evaluation of the momentum integrals from the temporal ones, for which so-called cutting rules (see sec.~5 below) offer a convenient book-keeping tool. Note also that many references use slightly differing conventions for the integration measures, with angular brackets often used to signify the fermionic nature of four-momenta.

\item A convenient simplification occurring in the strict $T=0$ limit is the exact vanishing of vacuum graphs containing no quark loops --- a direct consequence of the massless nature of gluons and the vanishing of scalefree integrals in dimensional regularization. At higher loop orders, the same mechanism leads to the vanishing of vacuum diagrams that contain quark loops but also a factorized purely bosonic sub-diagram, such as the first 11 diagrams of fig.~\ref{diags}.

\item A subtlety not present in studies of the short-lived quark-gluon plasma (QGP) created in heavy-ion collisions that needs to be taken into account when considering the high-density matter inside NSs is related to maintaining chemical (beta) equilibrium and local charge neutrality.\footnote{The reason one can neglect these effects in heavy-ion physics is related to the fact that the electromagnetic and weak interactions operate on much longer timescales than the strong-interaction processes.} To achieve both limits, one typically needs to add electrons to the system and implement a number of constraints between the chemical potentials of the relevant particle species. 

For the three lightest quark flavors present at NS densities, the requirement of beta equilibrium implies the relations $\mu_s=\mu_d$ and $\mu_u = \mu_d-\mu_e$ between the four chemical potentials present. This allows parametrizing the system in terms of only two chemical potentials, typically taken to be $\mu_d$ and $\mu_e$, while local charge neutrality adds one more constraint
\begin{eqnarray}
\frac{2}{3}n_u-\frac{1}{3}n_d-\frac{1}{3}n_s&=&n_e\, .
\end{eqnarray}
The last relation allows us to further solve $\mu_e$ as a function of $\mu_d$. 

In the simplifield limit of three massless quark flavors, typically a good approximation for bulk thermodynamic quantities at perturbative densities, the above equations admit the simple solution of $\mu_u=\mu_d=\mu_s$ and $\mu_e=0$. This means that both local charge neutrality and beta equilibrium can be maintained without the presence of electrons and with equal number densities for the three lightest quark flavors. Corrections to this limit due to a nonzero strange-quark mass can be  conveniently obtained using the quark-mass expansion scheme recently introduced in \cite{Gorda:2021gha}.
\end{itemize}

To get some hands-on experience on concrete perturbative calculations at high density, let us next consider the first two orders in the weak-coupling expansion of the pressure of cold and dense QM. Up to a $\mu$-independent vacuum part, the leading-order (LO) pressure of a system of noninteracting quarks at nonzero flavor-dependent quark chemical potentials $\mu_f$ but vanishing temperature can be written in the simple form (cf.~eq.~(7.43) of \cite{Laine:2016hma})
\begin{eqnarray}
p_\text{QCD}^\text{LO}(\{\mu_f\}) & = & 2N_c\sum_f \int_\textbf{p}\int_{-\infty}^\infty \frac{{\text d}p_0}{2\pi}\ln\big[(p_0+i\mu_f)^2+p^2+m_f^2\big]\, , 
\end{eqnarray}
where the  number of colors $N_c$ is kept unspecified for the sake of generality. 

It is convenient to begin the calculation from the integral over $p_0$. We could in principle perform it directly in the above logarithmic form, but a more straightforward strategy is to first carry out a differentiation with respect to $E_p^2 \equiv p^2+m_f^2$ and later integrate the result over the same parameter. To this end, we consider the integral
\begin{eqnarray}
\int_{-\infty}^\infty\frac{{\text d}p_0}{2\pi}\frac{1}{(p_0+i\mu_f)^2+E_p^2} &=&
\int_{-\infty}^\infty \frac{\text{d}p_0}{2\pi} \frac{1}{(p_0+i\mu_f-iE_p)(p_0+i\mu_f+iE_p)}  \nonumber \\
&=&\frac{\theta(E_p-\mu_f)}{2E_p}\, , \label{I1}
\end{eqnarray}
where we closed the integration contour over the upper halfplane. Integrating this expression with respect to $E_p^2$ from $E_p^2=\mu^2$, where the contributions to the pressure vanish on physical grounds (this represents the quark mass threshold), we easily obtain 
\begin{eqnarray}
\int_{-\infty}^\infty \frac{{\text d}p_0}{2\pi}\ln\big[(p_0+i\mu_f)^2+p^2+m_f^2\big]&=&(E_p-\mu^{ }_f)\theta(E_p-\mu^{ }_f)    
\end{eqnarray}
and further
\begin{eqnarray}
p_\text{QCD}^\text{LO}(\{\mu_f\})&=&2N_c \sum_f 
 \int_\mathbf{p} (E_p-\mu^{ }_f)\theta(E_p-\mu^{ }_f) \nonumber \\
 &=&2N_c \sum_f 
 \int_\mathbf{p} (\mu^{ }_f-E_p)\theta(\mu^{ }_f - E_p) \nonumber \\
  &\underset{m_f\to 0}{\to}& \frac{N_c}{12\pi^2} \sum_f \mu_f^4 \; .
\end{eqnarray}
Here, we have at the second equal sign discarded two terms --- proportional to the 1 in $\theta(E_p-\mu^{ }_f)=1-\theta(\mu^{ }_f-E_p)$ --- that either correspond to a $\mu=T=0$ vacuum contribution or vanish in dimensional regularization, and at the last stage finally proceeded to the massless limit where the spatial momentum integral trivializes. It is worth comparing the simplicity of this computation to the complications one encounters with its finite-$T$ counterpart, considered, e.g., in sec.~7, Appendix A of \cite{Laine:2016hma}.

Proceeding next to the Next-to-Leading Order (NLO) in the massless limit, we encounter one two-loop diagram with a quark loop, first considered in \cite{Freedman:1976ub}. Carrying out the Lorentz and color contractions, we straightforwardly obtain 
\begin{eqnarray}
-\frac{1}{2}\raisebox{-0.42\height}{\includegraphics[height=1.2cm]{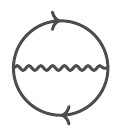}}&=&-d_A g^2\frac{d-1}{2}\int_{\widetilde{P}}\int_Q \frac{1}{P^2(P-Q)^2} \nonumber \\
&=& -d_A g^2\frac{d-1}{2}\bigg(\int_{\widetilde{P}}\frac{1}{P^2}\bigg)^2,
\end{eqnarray}
where $d_A=N_c^2-1$ is the dimension of the adjoint representation of the gauge group SU($N_c$) and we have explicitly included the negative sign and symmetry coefficient of the diagram on the left hand side of the equation. A clean factorization of the result into a product of two one-loop (fermionic) integrals can clearly be observed from the final form of the result. 

The sole one-loop master integral appearing in the above expression is clearly identical to the one in eq.~\ref{I1}, from where we can immediately take the result for the $p_0$ integration. A straightforward calculation utilizing this intermediate result produces now
\begin{eqnarray}
\int_{\widetilde{P}}\frac{1}{P^2} &=& -\frac{1}{4\pi^2}\int_0^\infty \text{d}p \, p \, \theta(\mu-p)  
\;=\; -\frac{\mu^2}{8\pi^2}, \label{I2}
\end{eqnarray}
where we have set $d=3$ and dropped a scalefree integral that vanishes in dimensional regularization. With this result, we obtain the pressure of cold and dense QM up to NLO,
\begin{eqnarray}
p_\text{QCD}^\text{NLO}(\{\mu_f\})&=&\bigg(\frac{N_c}{12\pi^2}  -  \frac{d_Ag^2}{64\pi^4}\bigg) \sum_f \mu_f^4 \, ,
\end{eqnarray}
a result first derived close to 50 years ago \cite{Freedman:1976ub}. 

\section{Soft effective theories for deconfined QCD matter}

At the three-loop order, performing the integrations for individual vacuum diagrams becomes technically considerably more demanding, with very few graphs displaying factorization to lower-order integrals (see \cite{Davydychev:2023jto} for recent general results concerning factorization). In addition, explicit computations performed both at nonzero and vanishing temperatures display uncancelled IR divergences in specific vacuum diagrams featuring single gluon propagators raised to powers higher than unity due to the presence of self-energy-type insertions. Unlike the UV divergences that cancel upon renormalization, these divergences are of a physical origin, related to contributions from long-distance interactions mediated by low-momentum gluons. To cancel them, something more is required, and here the discussion naturally separates to two different realms: nonzero and vanishing temperatures.

We begin the story from the perhaps more familiar limit of high temperatures, where the boundary conditions of the fields contributing to eq.~(\ref{path}) allow us to express their dependence on the imaginary time coordinate $x_0\equiv \tau$ through discrete Fourier series. In practice, we write
\begin{eqnarray}
\phi(\tau,\mathbf{x})&=&T\sum_{n=-\infty}^\infty \widetilde{\phi}_n(\mathbf{x})e^{i\omega^\text{bos}_n \tau}, \\
\psi(\tau,\mathbf{x})&=&T\sum_{n=-\infty}^\infty \widetilde{\psi}_n(\mathbf{x})e^{i\omega^\text{fer}_n \tau} \, ,
\end{eqnarray}
for bosonic and fermionic fields $\phi$ and $\psi$, respectively, with the Matsubara frequencies $\omega_n^\text{bos}=2n\pi T$ and $\omega_n^\text{fer}=(2n+1) \pi T$ acting as thermal mass terms for the three-dimensional field components carrying the index $n$. At high $T$, all three-dimensional fields except for the massless $n=0$ modes of bosons are protected againts IR problems by these nonvanishing masses, so that the minimal effective theory capable of describing the IR physics of hot QCD becomes a three-dimensional theory for the $n=0$ components of the $A_0$ and $A_i$ fields. What sets the Lorentz components of the original gauge field apart here is the fact that  the gauge field of a three-dimensional theory only has three (spatial) components. This makes the $n=0$ component of $A_0$ an adjoint scalar field in the effective theory, capable of acquiring a nonzero mass term in the corresponding effective Lagrangian.

The process of building three-dimensional effective field theories for the description of high-temperature QFTs by integrating out the nonzero Matsubara modes of four-dimensional fields is known as \textit{dimensional reduction}. It possesses a long history dating back to the original work of Appelquist and Pisarski in 1981 \cite{Appelquist:1981vg} and subsequent refinements in the 1990s by Kajantie et al.~\cite{Kajantie:1995dw} as well as Braaten and Nieto \cite{Braaten:1995cm}. For QCD, a particularly important step was taken in \cite{Braaten:1995jr}, where the widely-used terms Electrostatic and Magnetostatic QCD (EQCD and MQCD for short) were coined for the three-dimensional theories capable of describing the equilibrium physics of length scales $x\gtrsim 1/(gT)$ and $1/(g^2T)$, respectively. While the former of these theories takes the form of a three-dimensional Yang-Mills theory with an adjoint scalar (see, e.g., \cite{Laine:2016hma} for more details), 
\begin{eqnarray}
\mathcal{L}_\text{EQCD}&=&\frac{1}{4}F^a_{ij}F^a_{ij}+\text{tr}[D_i,A_0]^2+m_\text{E}^2\text{tr}A_0^2+\lambda_1\big(\text{tr}A_0^2\big)^2+\lambda_2\text{tr}A_0^4 \, ,\;\;\;
\end{eqnarray}
in MQCD the $A_0$ field with an $O(gT)$ mass is integrated out as well. These theories can be used to account for the soft contributions to equilibrium thermodynamic quantities such as the equation of state, but at lower temperatures their use is limited not only by the lack of a scale hierarchy between the hard ($\pi T$) and soft ($gT$) scales, but also by the explicit breaking of the Z($N_c$) center symmetry of the Yang-Mills part of QCD. This can be further remedied by supplementing the effective theory with a more versatile field content, thus extending the applicability of dimensional reduction to somewhat lower temperatures \cite{Vuorinen:2006nz,deForcrand:2008aw}.

\begin{figure}[t]
\centerline{%
\includegraphics[width=11cm]{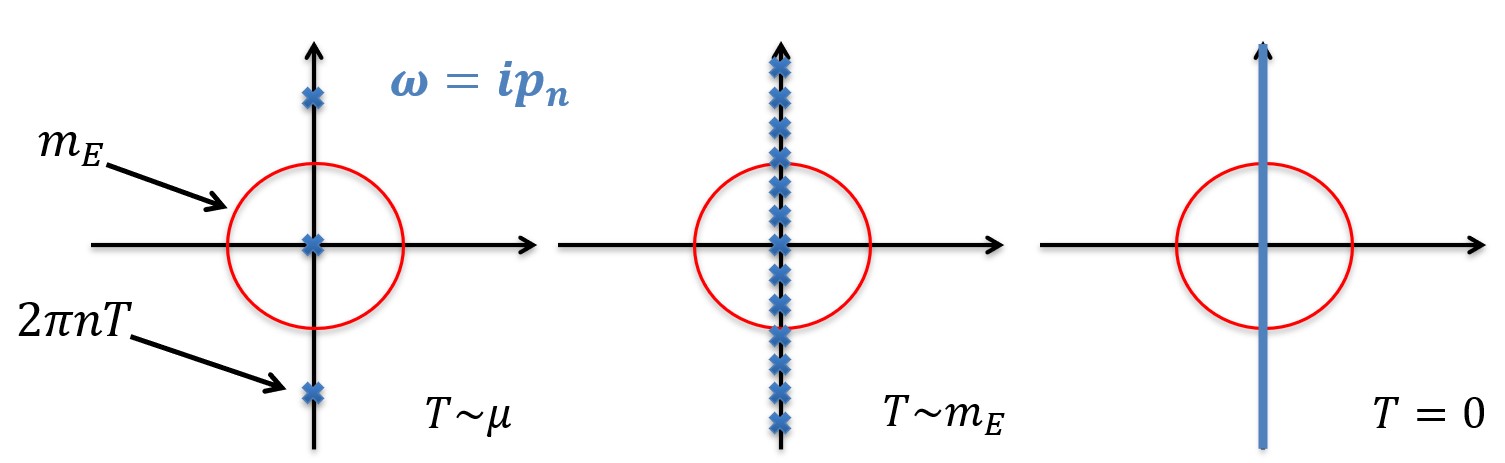}}
\caption{An illustration of how more and more bosonic Matsubara frequencies fit inside a circle of constant radius $m_\text{E}$ as the temperature is lowered first to a small number times $m_\text{E}\sim g\mu$ and then all the way to the $T=0$ limit. Note that for fermions, the result would be very different due to the imaginary offset of the Matsubara frequency by $i\mu$.}
\label{smallerT}
\end{figure}

The effects of quark chemical potentials can be introduced to the dimensionally reduced EFTs in a fairly straightforward manner \cite{Hart:2000ha,Vuorinen:2003fs}, with the main effects seen in small shifts of the EFT parameters from their $\mu=0$ values and in the introduction of one new EQCD operator of the form $\text{tr} A_0^3$. Upon increasing the parameter $\mu/T$ to values greatly exceeding unity, one, however, eventually exists the regime of validity of dimensional reduction. As demonstrated in \cite{Ipp:2006ij}, this takes place when the temperature is reduced below the scale $T\sim m_\text{E}$, where $m_\text{E}$ stands for the leading-order electric screening mass
\begin{eqnarray}
m_\text{E}^2&=& g^2\bigg[\frac{C_A+T_F}{3}T^2+\frac{1}{2\pi^2}\sum_f \mu_f^2\bigg] \label{mE}
\end{eqnarray}
and $T_F = N_f/2$. The physical effect that takes place when $T$ is lowered well below $m_\text{E}$ is that more and more bosonic Matsubara modes become soft and need to be resummed, a process depicted in fig.~\ref{smallerT}. In other words, the field content of EQCD is no longer sufficient to capture all of the soft physics in need of a resummed treatment.

The minimal effective theory replacing EQCD at low temperatures turns out to be a four-dimensional EFT constructed for all soft gluon modes satisfying $P^2\lesssim m_\text{E}^2$ and thus requiring some form of resummed treatment.\footnote{Note that fermions continue to be protected against IR problems also in the cold and dense limit through their nonvanishing chemical potentials. This can be understood in physical terms from the fact that all quantum-mechanical states are filled within the Fermi sphere, implying that only hard fermions with momenta of order $\mu$ can propagate.} This EFT, the form of which can be most straightforwardly derived by considering the soft-external-momentum limit of gluonic amplitudes, is dubbed Hard Thermal Loops (HTL) and was first introduced by Braaten and Pisarski already in 1989 \cite{Braaten:1989mz,Braaten:1991gm}. Since then, the HTL framework has been successfully applied to the derivation of numerous physical quantities including the bulk thermodynamic properties of high-temperature QGP at vanishing \cite{Andersen:1999fw} and nonzero chemical potentials \cite{Haque:2014rua}, the EoS of cold and dense QCD and QED \cite{Gorda:2021znl,Kurkela:2016was,Gorda:2022zyc,Gorda:2022fci,Gorda:2023mkk}, dynamical observables such as production and decay rates \cite{Braaten:1990wp,Ghiglieri:2013gia}, and even transport coefficients at various orders of perturbation theory \cite{Heiselberg:1993cr,Ghiglieri:2018dib}.

\section{Hard Thermal Loops: basic properties and simple applications}

The development of the HTL effective theory has been documented in multiple review articles and even textbooks, including e.g.~\cite{Ghiglieri:2020dpq,Bellac:2011kqa,Haque:2024gva}, and the theory itself has been extended to new physical realms and higher perturbative orders even relatively recently \cite{Caron-Huot:2007cma,Gorda:2023zwy}. To keep the present discussion at a tractable level, we will limit its treatment here to the parts of the theory necessary for determining the leading soft contributions to the pressure of cold and dense QM. This implies we will be mainly interested in the one-loop HTL gluon self energy, which characterizes how the propagation of soft gluons is modified by the medium and that can be obtained from a particular (soft-external-momentum) limit of the same quantity in full QCD. While the result can be shown to take a universal form, being equally applicable in the high-temperature regime, we will for simplicity perform the computation in the strict $T=0$ limit, allowing some technical simplifications to be implemented. After the derivation of the leading-order HTL self energies, we will apply the result to the determination of the leading non-analyticity in $\alpha_s$ of the weak-coupling expansion of the pressure of cold and dense QM.

The gluon self energy, or the gauge field self-energy tensor, is defined by the Schwinger-Dyson equation, which relates it to  the difference of the full and bare inverse propagators,
\begin{eqnarray}
\Pi_{\mu\nu}^{ab}(P)&=&\big(D^{-1}\big)_{\mu\nu}^{ab}(P) - \big(D_{(0)}^{-1}\big)_{\mu\nu}^{ab}(P)\, .
\end{eqnarray}
Here, the free propagator takes the usual form
\begin{eqnarray}
\big(D_{(0)}\big)_{\mu\nu}^{ab}(P)&=&    \frac{\delta_{\mu\mu}-(1-\xi)P_\mu P_\nu/P^2}{P^2}\delta^{ab}
\end{eqnarray}
in the covariant gauges, with $\xi$ being the corresponding gauge parameter. Simply put, $\Pi_{\mu\nu}^{ab}(P)$ is thus gauge-theory generalization of the scalar-field self energy, which dresses the massless scalar propagator as $\frac{1}{P^2}\to \frac{1}{P^2+\Pi(P)}$.

Although a symmetric rank two tensor in four dimensions in principle contains 10 independent components, various symmetries of the system can be seen to significantly reduce this number. First, while four-dimensional Lorentz invariance is broken by the existence of a preferred frame in a system in thermal equilibrium --- the rest frame of the heat bath of the medium --- three-dimensional rotational invariance is enough to strongly limit the possible structures appearing in the result. In practice, it implies that the self-energy tensor can be given as a linear combination of at most four independent elements: the metric $\delta_{\mu\nu}$ and three symmetric rank-two tensors composed with the external momentum $P_\mu$ and the four-vector singling out the temporal direction, $n_\mu\equiv \delta_{\mu 0}$. Current conservation, implemented via the Slavnov-Taylor identities, further restricts the possible form of the tensor by requiring it to be transverse with respect to the external four-momentum in vacuum. As discussed in detail in \cite{Weldon:1996kb,Gorda:2023zwy}, the situation becomes more complicataed in a thermal setting, where transversality depends on whether non-Abelian interactions are present, what loop order is being considered, and even which (covariant) gauge has been chosen. Here, it suffices to note that for QCD at nonzero $\mu$ but vanishing temperature, the one-loop self energy must be transverse irrespective of the gauge one is working in.

The transversality of the LO self energy implies that the tensor can possess at most two independent components, typically labeled according to their transversality properties with respect to the external three-momentum. Omitting the Kronecker delta function in the adjoint color indices, the result of this decomposition reads
\begin{eqnarray}
\Pi_{\mu\nu}(P)&=& \Pi^\text{vac}_{\mu\nu}(P)+\Pi_\text{T}(P)\mathbb{P}_{\mu\nu}^\text{T}(P)+\Pi_\text{E}(P)\mathbb{P}_{\mu\nu}^\text{E}(P)\, , \label{Pimunu}
\end{eqnarray}
where the UV divergent vacuum part $\Pi^\text{vac}_{\mu\nu}(P)$ represents the $T\to 0$, $\mu\to 0$ limit of the polarization tensor and can be seen to be proportional to $P^2\delta_{\mu\nu}-P_{\mu}P_{\nu}$, being thus subdominant for soft external momenta. The matter part of the self energy is on the other hand divided into a transverse (T) and a longitudinal or Euclidean (E) part, with the corresponding projection operators reading
\begin{eqnarray}
\mathbb{P}_{\mu\nu}^\text{T}(P)&\equiv&\delta_{\mu i} \delta_{\nu j}\Big(\delta_{ij}-\frac{p_i p_j}{p^2}\Big) \, , \\    
\mathbb{P}_{\mu\nu}^\text{E}(P)&\equiv&\delta_{\mu \nu}-\frac{P_{\mu} P_{\nu}}{P^2}-\mathbb{P}_{\mu\nu}^\text{T}(P) \, .
\end{eqnarray}
It is straightforward to verify that apart from the obvious transversality of $\mathbb{P}_{\mu\nu}^\text{T}(P)$ with respect to the external three-momentum, these tensors satisfy the following properties (see, e.g., \cite{Kapusta:2006pm}):
\begin{eqnarray}
\mathbb{P}_{\alpha\beta}^\text{X}(P)\mathbb{P}_{\beta\gamma}^\text{X}(P)&=&\mathbb{P}_{\alpha\gamma}^\text{X}(P)\, , \quad \text{X}\;=\;\text{T,\, E}\,, \\
\mathbb{P}_{\alpha\beta}^\text{X}(P)\mathbb{P}_{\beta\gamma}^\text{Y}(P)&=&0\, , \quad \text{X}\;\neq\;\text{Y}\,, \\
\mathbb{P}_{\mu\mu}^\text{T}(P)&=& d-1\, , \quad \mathbb{P}_{\mu\mu}^\text{E}(P)\;=\; 1 \, .
\end{eqnarray}

Moving on to the actual evaluation of the gluon self energy, we first note that the one-loop gluon polarization tensor can be obtained from the sum of all amputated one-particle-irreducible Feynman diagrams with two external gluon lines. Its general expression at nonzero $T$ and $\mu$ is rather lengthy, and we refer the interested reader to eq.~(C1) of \cite{Ipp:2006ij}, where the Feynman gauge version of this quantity is displayed. In the strict $T=0$ limit, some simplifications occur, though, and e.g.~the purely bosonic term only contributes to the vacuum limit of the tensor. If we are moreover interested in the soft (or HTL) limit of the result, where $|P| \ll \mu$ and the components of the loop momentum $Q$ can subsequently be assumed to be dominant over those of $P$, we observe part of the fermionic contribution dropping out as well. This leaves us with the much more manageable $T=0$ expression
\begin{eqnarray}
\Pi_{\mu\nu}(P)|_{T=0}&=& -g^2\sum_{f}\bigg\{\int_{\widetilde{Q}} \frac{2\delta_{\mu\nu}}{Q^2} - \int_{\widetilde{Q}} \frac{(2Q-P)_\mu (2Q-P)_\nu}{Q^2(Q-P)^2} \bigg\} 
\end{eqnarray}
to work with, where the fermionic integration measures remind us of the presence of $\mu_f$ in the corresponding fermion propagators.

To derive results for the two functions $\Pi_\text{T}(P)$ and $\Pi_\text{E}(P)$, it suffices to consider two independent components or contractions of the self energy tensor, for which we choose $\Pi_{\mu\mu}$ and $\Pi_{00}$. Omitting terms that are clearlt subdominant for soft external momentum $P$, we obtain
\begin{eqnarray}
\Pi_{\mu\mu}(P)|_{T=0}&\approx& -2(D-2)g^2\sum_{f}\int_{\widetilde{Q}} \frac{1}{Q^2} \, , \label{Pimumu} \\
\Pi_{00}(P)|_{T=0}&\approx& -g^2\sum_{f}\bigg\{\int_{\widetilde{Q}} \frac{2}{Q^2} - \int_{\widetilde{Q}} \frac{(2q_0-p_0)^2}{Q^2(Q-P)^2} \bigg\} \, , \label{Pi00}
\end{eqnarray}
of which we have already evaluated the tadpole integral in eq.~(\ref{I2}). To this end, let us here concentrate on the latter term in the above expression for $\Pi_{00}$, i.e.~the only integral with nontrivial dependence on $P$.

A key observation simplifying the evaluation of the $q_0$ integral in
\begin{eqnarray}
\widetilde{\Pi}(P)&\equiv&\int_{\widetilde{Q}} \frac{(2q_0-p_0)^2}{Q^2(Q-P)^2}
\end{eqnarray}
is that two of the four poles, including  $q_0=-i\mu-iq$, always reside on the lower halfplane, while two, including  $q_0=-i\mu+iq$, can reside on either halfplane depending on the magnitude of the three-momentum $q$ in comparison with the chemical potential $\mu$. Writing the resulting step function in the form $\theta(q-\mu)=1-\theta(\mu-q)$ and recognizing the appearance of a $\mu$-independent integral that can be dropped, we straightforwardly obtain for the vacuum-subtracted "matter" part of the function
\begin{eqnarray}
\widetilde{\Pi}(P)_\text{mat}&=& -\frac{1}{2}\int_q\frac{\theta(\mu-q)}{q} \bigg\{\frac{(2iq-p_0)^2}{(iq-p_0)^2+(\mathbf{q}-\mathbf{p})^2}+\frac{(2iq+p_0)^2}{(iq+p_0)^2+(\mathbf{q}-\mathbf{p})^2}\bigg\} \nonumber \\
&=& -\frac{1}{8\pi^2}\int_0^\mu \text{d}q q \int_0^\pi \text{d}\theta \sin \theta \bigg\{\frac{(2iq-p_0)^2}{P^2-2iqp_0-2\mathbf{q}\cdot\mathbf{p}}+\text{c.c.}\bigg\} \, , \label{Pitilde}
\end{eqnarray}
where we have on the latter row set $d=3$ and abbreviated the complex conjugate of the first expression (for real $P_\mu$) by c.c. 

The two integrations remaining in eq.~(\ref{Pitilde}) can now be straightforwardly carried out. Upon an expansion of the result to leading order in in $P/q$, the final expression becomes
\begin{eqnarray}
\widetilde{\Pi}(P)_\text{mat}^\text{soft}&=& \frac{\mu^2}{4\pi^2}\bigg\{1-\frac{ip_0}{p}\ln\frac{ip_0+p}{ip_0-p} \bigg\} \, ,
\end{eqnarray}
implying that we have obtained a remarkably simple analytic form for the soft-external-momentum limit of the original integral. In particular, all dependence on the hard scale $\mu$ has neatly factorized from the rest, leaving behind a nontrivial function of the ratio of the two independent components of the external momentum, $p_0$ and $p=|\mathbf{p}|$.

Combining the above result with that of  eq.~(\ref{I2}), we are now ready to write down final expressions for the HTL self energy components $\Pi_\text{T}(P)$ and $\Pi_\text{E}(P)$. Setting $d=3$ everywhere, a simple calculation produces
\begin{eqnarray}
\Pi_\text{E}(P)&=&\frac{P^2}{p^2}\Pi_{00}^\text{mat}(P)\,, \quad \Pi_\text{T}(P)\;=\;\frac{1}{2}\Pi_{\mu\mu}^\text{mat}(P)-\frac{P^2}{2p^2}\Pi_{00}^\text{mat}(P) \, ,
\end{eqnarray}
using which [and taking into account the sum over flavors in eqs.~(\ref{Pimumu}) and (\ref{Pi00})] we quickly reach the final gauge-invariant result
\begin{eqnarray}
 \Pi^\text{HTL}_\text{T}(P) &=& 
 -\frac{m_\text{E}^2}{2}\frac{P^2}{p^2} \bigg\{\frac{p_0^2}{P^2}-  
  \frac{i p^{ }_0}{2 p}
  \ln \frac{ip^{ }_0 + p}{ip^{ }_0 - p} 
 \bigg\}
  \,, \label{HTLT}  \\
 \Pi^\text{HTL}_\text{E}(P) &=&
 m_\text{E}^2\frac{P^2}{p^2}
  \bigg[ 
    1 -  \frac{i p^{ }_0}{2 p} 
    \ln \frac{ip^{ }_0 + p}{ip^{ }_0 - p} 
  \bigg]
  \, . \label{HTLE}
\end{eqnarray}
with $m_\text{E}$ denoting the leading-order electric screening mass from eq.~(\ref{mE}). With these self-energies, the full HTL-resummed gluon propagator finally becomes
\begin{eqnarray}
D_{\mu\nu}^{ab}(P)&=&  \delta^{ab}\bigg\{\frac{\mathbb{P}_{\mu\nu}^\text{T}(P)}{P^2+\Pi_\text{T}(P)}+\frac{\mathbb{P}_{\mu\nu}^\text{E}(P)}{P^2+\Pi_\text{E}(P)}+\frac{\xi P_\mu P_\nu}{(P^2)^2}\bigg\} \, ,\end{eqnarray}
where we have again reinstated the gauge-parameter term from the free propagator.

Interestingly, the final result for the HTL self energies and the HTL-resummed gluon propagator take precisely the same forms as in the more general case of nonzero temperatures and chemical potentials; see, e.g., sec.~8 of \cite{Laine:2016hma}. It is also worth noting that both self-energy components depend on the external momentum only through the ratio $\frac{p_0}{p}$, so that defining an angle $\phi$ via $\tan\phi\equiv p/p_0$ we can parameterize the entire LO HTL self energy as a function of this single dimensionless variable. This result suffices for typical low-order computations utilizing the HTL effective theory, while at higher orders, one may, depending on the quantity being determined, also require the use of the HTL vertex functions, the HTL fermion self energy, or subleading corrections to the above result from higher loop orders or a power expansion in $P/\mu$. For detailed derivations of these additional features of the HTL theory, we recommend a recent review by Haque and Mustafa \cite{Haque:2024gva} as well as the original research article \cite{Gorda:2023zwy}, where the real-time formalism of perturbation theory is applied to the gluon polarization tensor at nonzero $T$ and $\mu$.

As a straightforward application of the LO HTL self energy derived above, let us now briefly return to the weak-coupling expansion of the pressure of dense zero-temperature QM. Here, we first note that the only three-loop vacuum diagram displaying an IR divergence in the $T=0$ limit can be written in the form
\begin{eqnarray}
\frac{1}{4}\raisebox{-0.42\height}{\includegraphics[height=1.2cm]{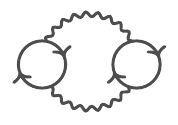}}&=& \frac{d_A}{4}
 \int_P\bigg\{
  (d-1)\frac{\Pi^{2}_\text{T}(P)}{(P^2)^2}
 +\frac{\Pi^{2}_\text{E}(P)}{(P^2)^2}+\cdots
  \bigg\}  \label{3loopIR} \\
  &\to& \frac{d_A}{4}
 \int_P\bigg\{
  (d-1)\frac{\big(\Pi^\text{HTL}_\text{T}(P)\big)^2}{(P^2)^2}
 +\frac{\big(\Pi^\text{HTL}_\text{E}(P)\big)^2}{(P^2)^2}
  \bigg\} \, , \;\;\; \nonumber
\end{eqnarray}
where we first utilized the expansion (\ref{Pimunu}) for the gluon self energy, left out IR-convergent terms generated by the vacuum part, and finally retained only the LO IR behavior of the integrand by replacing the full self energies by their HTL limits. 

To cure the IR divergence in the final form of eq.~(\ref{3loopIR}), the simplest way to proceed is to explicitly resum all ring diagrams of the same type, each taking the form  \begin{eqnarray}
I_\text{ring}^N &\equiv& \frac{(-1)^N d_A}{2N}
 \int_P\bigg\{
  (d-1)\frac{\big(\Pi^\text{HTL}_\text{T}(P)\big)^N}{(P^2)^N}
 +\frac{\big(\Pi^\text{HTL}_\text{E}(P)\big)^N}{(P^2)^N}
  \bigg\}\, ,
\end{eqnarray}
with $N$ denoting the number of one-loop self-energy insertions in the graph. Recognizing here the coefficients of the Taylor expansion of the function $\ln (1+x)$, we may explicitly perform the sum $\sum_{N=1}^\infty I_\text{ring}^N$, obtaining a quantity identifiable as the LO pressure of the HTL effective theory,
\begin{eqnarray}
 p_\text{HTL}^\text{LO}&=&-\frac{d_A}{2}
 \int_P\Big\{
  (d-1)\ln\big[ K^2+ \Pi^\text{HTL}_\text{T}(P)\big]
 +\ln\big[ K^2+\Pi^\text{HTL}_\text{E}(P) \big]
  \Big\}  \nonumber \\
&=&-\frac{d_A m_\text{E}^{D}}{2}
 \int_{\hat P}  
 \Big\{ (
 d-1)\ln\big[ \hat{P}^2 + \hat{\Pi}^{ }_\text{T}(\phi)\big]
 +\ln\big[ \hat{P}^2+\hat{\Pi}^{ }_\text{E}(\phi)\big]
 \big\} \, . \label{pHTLLO} 
\end{eqnarray}
Here, we have on the lower line defined the scaled quantities $P\equiv m_\text{E}^{ } {\hat P}$ and $\Pi^\text{HTL}_\text{X}(P)\equiv m_\text{E}^2\,{\hat \Pi}_\text{X}(\phi)$, with X standing for either T or E. 

To proceed from here, we first make use of the standard integral
\begin{eqnarray}
 \int\! \frac{{\text d}^D K}{(2\pi)^D}
 \ln(K^2+m^2) &=& 
  - \frac{m^D_{ }\Gamma(-D/2)  }{(4\pi)^{\frac{D}{2}}}
\end{eqnarray}
that can be used to take care of the radial integration in eq.~(\ref{pHTLLO}). This quickly produces (see \cite{Laine:2016hma} for details)
\begin{eqnarray}
 p_\text{HTL}^\text{LO}&=&
 \frac{d_Am_\text{E}^{D} \,\Gamma(D/2)\Gamma(-D/2)}
      {(4\pi)^{\frac{D+1}{2}}\Gamma((D-1)/2)} \times \nonumber \\
      &\times&
 \int_0^\pi{\text d}\phi\,\sin^{2-2\epsilon}\!\phi 
 \Big[(d-1)\,\hat{\Pi}^{D/2}_\text{T}(\phi)
 +\hat{\Pi}^{D/2}_\text{E}(\phi)\Big]\, , \;\;\;\; \label{pHTLLO2}
\end{eqnarray}
where the prefactor is of order $m_\text{E}^{4-2\epsilon}/\epsilon$. It is precisely this product of a $D$-dependent power of $m_\text{E}$ and $1/\epsilon$ that upon a power expansion in $\epsilon$ gives rise to a term of type $\alpha_s^2\ln\alpha_s$ in the weak-coupling expansion of the QM pressure. 

Setting now $\epsilon=0$ in the angular integral of eq.~(\ref{pHTLLO2}), a straightforward calculation produces
\begin{eqnarray}
 \int_0^\pi{\text d}\phi\,\sin^{2}\!\phi 
 \Big[(d-1)\,\hat{\Pi}^{2}_\text{T}(\phi)+\hat{\Pi}^{2}_\text{E}(\phi)
 \Big]&=&\frac{\pi}{4}\, .
\end{eqnarray}
With this result we may finally read off the coefficient of the leading nonanalyticity in the weak-coupling expansion of the pressure, 
\begin{eqnarray}
  p_\text{HTL}^\text{LO}&=&
  \frac{d_A m_\text{E}^4\, \Lambda^{-2\epsilon}}{8(4\pi)^2}
  \bigg(\frac{\Lambda^2}{m_\text{E}^2}\bigg)^\epsilon_{ }
 \frac{1}{\epsilon}+O(\epsilon^0) = -\frac{d_A m_\text{E}^4}{8(4\pi)^2}\ln\alpha_s +\cdots \, ,
\end{eqnarray}
a  contribution first derived by Freedman and McLerran in 1977 using considerably more involved computational techniques \cite{Freedman:1976ub}. Note that the corresponding $1/\epsilon$ divergence that we have discarded here is of UV type (in the HTL effective theory) and cancels against a corresponding IR divergence in the three-loop full theory vacuum diagram considered in eq.~(\ref{3loopIR}).

\section{QM thermodynamics at three and four loops}

Next, we move on to the full three-loop order and beyond, i.e.~towards the state of the art in perturbative calculations for QCD thermodynamics. Up to and including $O(g^5)$ contributions, the pressure of deconfined QCD matter at nonzero temperature and density is most conveniently expressed in a form first introduced in \cite{Kurkela:2016was},
\begin{eqnarray}
p_\text{QCD}&=& p_\text{QCD}-p_\text{soft} + p_\text{soft} \nonumber \\
&=& p_\text{QCD}^\text{naive}-p_\text{soft}^\text{naive} + p_\text{soft}^\text{res} \nonumber \\
&=&p_\text{QCD}^\text{naive}+p_\text{DR}^\text{res}-p_\text{DR}^\text{naive}+p_\text{HTL}^\text{res}-p_\text{HTL}^\text{naive}\, . \label{pressT}
\end{eqnarray}
Here, we have first added and subtracted the pressure of a properly defined soft effective theory from the full QCD pressure, then noticed that the difference $p_\text{QCD}-p_\text{soft}$ is an IR safe quantity that can be evaluated in a naive loop expansion in the respective theories, and finally introduced a dimensionally reduced (DR) effective theory as the minimal EFT for the $n=0$ Matsubara modes of gluons and HTL for all the non-static modes. 

The five terms appearing in eq.~(\ref{pressT}) are defined as follows:
\begin{itemize}
\item $p_\text{QCD}^\text{naive}$ denotes the naive loop expansion of the pressure of full QCD, evaluated up to and including the three-loop order. Dimensional regularization is used to regulate both UV and IR divergences, of which the former cancel upon renormalization. IR divergences of the form $1/\epsilon$, however, remain, in addition to which the $O(\alpha_s^2)$ part of the result contains terms that diverge as $\mu^4\ln \frac{T}{\mu}$ in the small-$T$ limit.
\item $p_\text{DR}^\text{res}$ denotes the pressure of the dimensionally reduced effective theory EQCD, evaluated in a weak-coupling expansion within this theory featuring a massive $A_0$ propagator. The resulting contribution to the pressure is IR safe by construction but contains a UV divergence.
\item $p_\text{DR}^\text{naive}$ denotes a version of the EQCD pressure evaluated by treating the $A_0$ mass as an interaction. This leads to an expression that identically vanishes in dimensional regularization due to the integrals becoming scalefree, but the calculation can be seen to contain a cancellation between equal but opposite UV and IR $1/\epsilon$ poles that convert the UV divergence of $p_\text{DR}^\text{res}$ into an IR one in the difference $p_\text{DR}^\text{res}-p_\text{DR}^\text{naive}$. This remaining IR divergence cancels against a similar term in $p_\text{QCD}^\text{naive}$.
\item $p_\text{HTL}^\text{res}$ and $p_\text{HTL}^\text{naive}$ denote the logarithmic HTL ring sum [cf.~eq.~(\ref{pHTLLO})] and its expansion in self-energies up to and including quadratic order, but with the contribution of the Matsubara zero mode left out. Both expressions are IR finite at nonzero $T$, but contain UV divergences that cancel  in the difference of the two terms. In the $T\to 0$ limit, the difference $p_\text{HTL}^\text{res}-p_\text{HTL}^\text{naive}$ develops an IR divergence proportional to $\ln T$, which is seen to cancel against the corresponding term in the $T\to 0$ limit of $p_\text{QCD}^\text{naive}$.
\end{itemize}
All in all, combining the five terms in the final form of eq.~(\ref{pressT}), we recover an expression that is free of both UV and IR divergences and agrees with all previously known limits of the pressure up to and including order $\alpha_s^{5/2}$. This was the main result of \cite{Kurkela:2016was} and continues to represent the state-of-the-art result for the pressure up to and including all fully known perturbative orders (i.e., not counting logarithms). The only exception to this is the large-$N_f$ limit of QCD, where an all-orders result for the high-temperature pressure has been determined thanks to simplifications that occur in precisely this limit (see, e.g., \cite{Moore:2002md,Ipp:2003zr,Ipp:2003yz,Gynther:2009qf}).

Proceeding beyond the three-loop order, the derivation of new terms in the weak-coupling expansion of the pressure becomes considerably more involved and the methods used in the low- and high-temperature regimes become increasingly disjoint. Here, we restrict our discussion to the strict $T=0$ limit, which was at the center of our attention already in the previous sections of these notes and which is importantly free of the famous Linde problem \cite{Linde:1980ts}, plaguing high-order perturbative calculations at finite temperature. Given the complexity of the problem and the fact that parts of it remain under active work, we will keep our discussion at a mostly qualitative level, concentrating on the structure of the weak-coupling expansion, the physical origins of various contributions entering at the $\alpha_s^3$ order, and the computational tools required in these calculations. For a reader interested in further computational details, we will provide numerous references to the original research articles below.

Up to and including the four lowest orders, the weak-coupling expansion of the pressure of cold ($T=0$) and dense unpaired QM matter obtains the schematic form  \cite{Gorda:2021znl,Gorda:2021kme}
\begin{eqnarray}
    p  = p_\text{FD} 
    +\alpha_s p_1^h   +\alpha_s^2 p_2^h  &+& \alpha_s^3 p_3^h \nonumber \\
     +\alpha_s^2 p_2^s&+&\alpha_s^3 p_3^s  \nonumber  \\
    &+& \alpha_s^3 p_3^m \, , \label{presschem}
\end{eqnarray}
where the coefficients $p_2^s$ and $p_3^m$ contain linear logarithms of the coupling $\alpha_s$ and the coefficient $p_3^s$ both linear and quadratic ($\ln^2\alpha_s$) logs \cite{Gorda:2018gpy}.\footnote{Note that many terms here also depend on logs of the renormalization scale $\bar{\Lambda}$ that cancel the scale dependence from the running of $\alpha_s$ order by order. We will mostly skip this topic here, but a reader interested in the precise mechanism, in which these cancellations occur, is referred to the Supplemental Material of \cite{Gorda:2022zyc}.} The letters $h$, $s$ and $m$ refer here to the "hard", "soft" and "mixed" sectors, corresponding to different regions of momentum space that contribute to the pressure at these orders. A detailed account of the different contributions can be found from \cite{Gorda:2021kme}, the results of which we summarize here, paying particular attention on the state-of-the-art $\alpha_s^3$ order.

\textbf{Hard sector:} Similarly to the case of the finite-temperature pressure in eq.~(\ref{pressT}), the contribution of the hard momentum scale $\mu_B$ to the pressure of cold and dense QM originates through a naive loop expansion of the full theory pressure. This necessitates the evaluation of all four-loop vacuum, or bubble, diagrams of QCD at nonzero chemical potentials but vanishing temperature, using dimensional regularization to regulate all divergences. The UV divergences of the result are expected to be removed upon renormalization, while --- similarly to the finite-$T$ calculation described above --- IR divergences will be seen to cancel against corresponding UV divergences from the EFT calculations corresponding to the soft and mixed contributions to the pressure.

At one-, two- and three-loop orders, the evaluation of vacuum diagrams can be performed analytically for vanishing quark masses, for which it was in fact completed for arbitrary values of $T$ and $\mu$ in \cite{Vuorinen:2003fs}. When a nonzero strange-quark mass is implemented, numerical methods are on the other hand required even in the strict $T=0$ limit, where the three-loop pressure was first determined in \cite{Kurkela:2009gj}. This computation relied on the use of the so-called cutting rules, where the underlying idea is to first perform all fermionic (but not bosonic) temporal momentum integrals, reducing the evaluation of the original Feynman integral to a sum of three-dimensional "phase-space" integrations over vacuum ($T=\mu=0$) on-shell amplitudes. Below, we we will briefly specify the cutting rules and apply them to a simple example calculation at the two-loop level, refering a reader interested in details of their formal derivation to the original article \cite{Ghisoiu:2016swa}. 

The cutting rules act on a scalarized $T=0$ Feynman integral corresponding to either a vacuum diagram or a Euclidean $N$-point function with real-valued external momenta. The integral is assumed to be composed of two types of (massive or massless) propagators, bosonic and fermionic, of which the former are composed of real-valued momenta while the temporal components of the latter are shifted by $i\mu$ as discussed in sec.~2 above. We furthermore assume that no fermionic propagator is raised to a power higher than unity, although this restriction can be relaxed if proper care is taken in the calculation \cite{Gorda:2022yex,Osterman:2023tnt}. 

With these assumptions, the cutting rules amount to the following set of five operations:
\begin{enumerate}[label=(\roman*)]
\item  Graphically perform all possible cuts of \textit{independent} internal fermion lines in the diagram, ranging from zero to the number of loops in the graph. Here, the independence of a given set of internal lines is equivalent to being able to choose the corresponding momenta as the loop momenta in the Feynman integral in question.
\item For each cut fermion line with momentum $P$, remove the corresponding propagator from the integral and place the same momentum on shell in the thus generated amplitude by writing $p_0=iE_p$, $E_p\equiv{p^2+m^2}$.
\item Set $\mu=0$ in each of the uncut fermion propagators and evaluate the corresponding integrals while assuming all components of the external momenta to be real-valued.
\item Integrate over the three-momenta of the cut lines with integration measure $-\theta(\mu-E_k)/(2E_k)$ (times the usual $1/(2\pi)^3$ associated with momentum integrals), with the integrand being the vacuum ($\mu=0$) on-shell amplitude generated by the cutting.
\item Sum all the individual contributions together.
\end{enumerate}
This procedure reduces the evaluation of the original finite-$\mu$ diagram to a number of separate terms, in which the $\mu$ dependence resides only in the theta functions of the phase space integration measures. This is a significant simplification, as the values of the vacuum amplitudes can typically be taken from collider-physics literature and the phase space integrals are by construction UV finite.

\begin{figure}[t]
\centerline{%
\includegraphics[width=10cm]{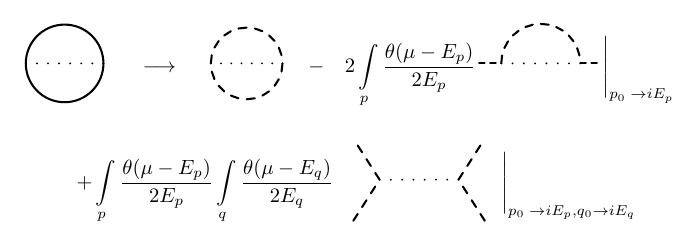}}
\caption{An graphical illustration of the cutting of the two-loop graph $I_2(\mu)$ discussed in the main text, with solid lines corresponding to a massive scalarized fermion propagator with nonzero $\mu$, dashed lines to its $\mu=0$ version, and dotted lines to a massless bosonic propagator. The figure is taken from \cite{Ghiglieri:2020dpq}.}
\label{cuts}
\end{figure}

As a simple example case, we inspect next the $T=0$, $\mu\neq 0$ integral
\begin{eqnarray}
I_2(\mu)&=&\int_{\widetilde{P}} \int_{\widetilde{Q}} \frac{1}{P^2+m^2}\frac{1}{Q^2+m^2}\frac{1}{(P-Q)^2} \;
\end{eqnarray}
where the momenta $P$ and $Q$ correspond to massive fermions at nonzero chemical potential and $P-Q$ to a massless boson. Identifying two as the number of independent fermionic momenta, the cutting rules amount to writing the integral in the form (see fig.~\ref{cuts})
\begin{eqnarray}
I_2(\mu)&=& I_2^\text{0-cut}+ I_2^\text{1-cut}(\mu)+I_2^\text{2-cut}(\mu)\,,
\end{eqnarray}
where the 0-cut part refers to the (uninteresting) $\mu=0$ version of the original integral, while the 1- and 2-cut parts carry all the  $\mu$-dependence of the original graph:
\begin{eqnarray}
I_2^\text{1-cut}(\mu)&=& -2\int_p \frac{\theta(\mu-E_p)}{2E_p} \Bigg[\int_Q \frac{1}{Q^2+m^2}\frac{1}{(P-Q)^2}\Bigg]_{p_0\to iE_p},\\
I_2^\text{2-cut}(\mu)&=& \int_p \frac{\theta(\mu-E_p)}{2E_p}\int_q \frac{\theta(\mu-E_q)}{2E_q} \Bigg[\frac{1}{(P-Q)^2}\Bigg]_{p_0\to iE_p,\,q_0\to iE_q}\, . \nonumber
\end{eqnarray}
The resulting integrals can be performed numerically at ease, or even analytically for $m=0$ (see appendix D of \cite{Osterman:2023tnt} for details), to reach the final result for the two-loop diagram. Note that the factor of two in the one-cut expression above results from the two separate one-cut terms producing identical results due to a symmetry of the original integral. It should be noted that such symmetries are not always present in higher-loop cases, and care must be taken to not automatically equate two terms that differ, e.g., by the relative direction of fermion flow in two separate fermion loops.

\begin{figure}[t]
\centerline{%
\includegraphics[width=\textwidth]{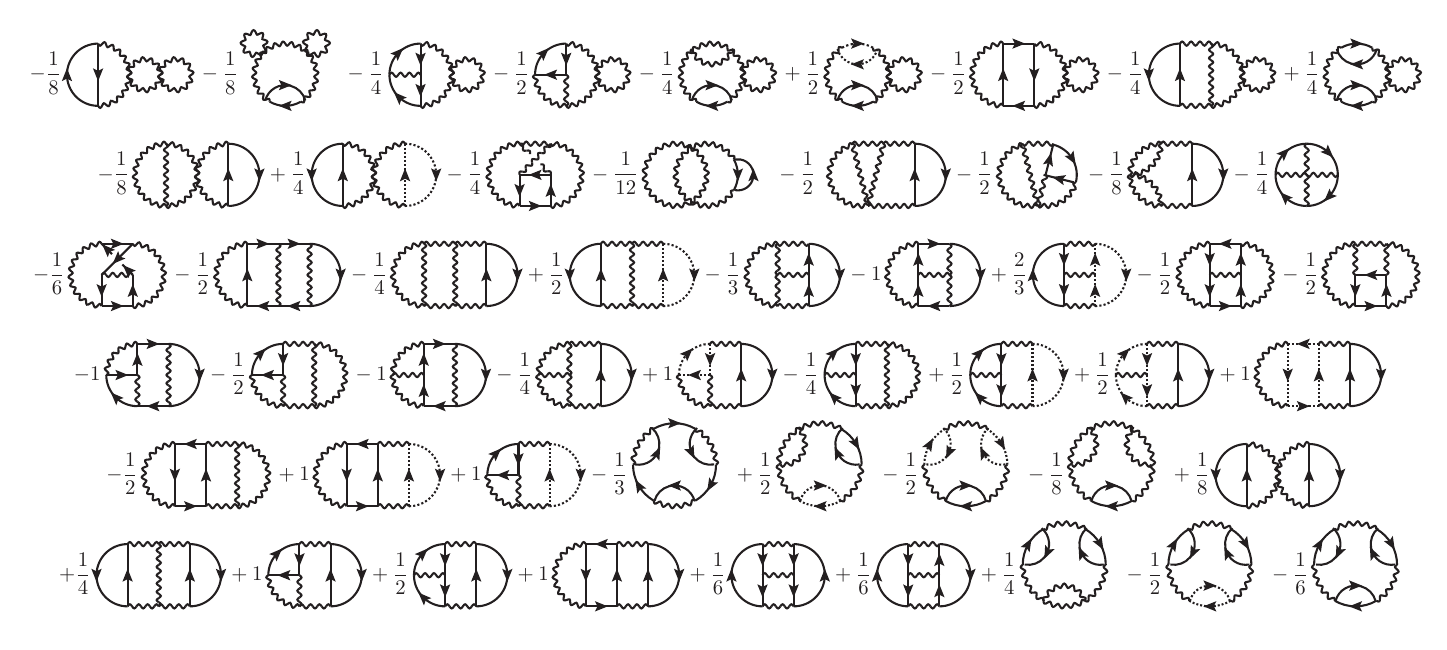}}
\caption{A list of all four-loop vacuum diagrams of QCD containing at least one fermion loop, shown together with the associated signs and symmetry coefficients. The first 12 diagrams can be shown to exactly vanish due to either containing a factorized scalefree integral (first 11 graphs) or being proportional to a vanishing color algebra (graph 12). The figure has been composed by Pablo Navarrete.}
\label{diags}
\end{figure}

The cutting rules were first used in \cite{Kurkela:2009gj} to evaluate all massive three-loop vacuum diagrams of QCD in a calculation that would have been considerably more challenging without this crucial computational aid. Proceeding all the way to four loops, the complexity of the calculations becomes considerably more challenging, however, and only two out of the altogether 52 individual graphs depicted in fig.~\ref{diags} have been fully evaluated so far \cite{Gorda:2022fci,Navarrete:2024zgz}. From the rest, the first 12 of fig.~\ref{diags} are found to vanish because they either contain a factorized scalefree integral or their color trace gives zero. Active work is currently underway to evaluate the remaining diagrams, with promising first steps taken recently in a yet unpublished article \cite{Aapeli}. The steps involved in this process include performing the color traces and Lorentz algebra of all graphs (in a general covariant gauge), then dividing the result into group-theory-invariant sectors, and finally systematically implementing momentum shifts and other standard manipulations to achieve the cancellation of the covariant gauge parameter $\xi$ within each group theory sector. 

After the above process, we are left with a relatively high number of independent master integrals (of order 100) to evaluate, which one may try to reduce using various novel methods, including but not limited to four-dimensional IBP identities recently generalized to the $T=0$, $\mu\neq$ realm in \cite{Osterman:2023tnt}. Finally, once the seemingly irreducible masters have been identified, the remaining task is to evaluate them starting from the ones that factorize into lower-loop-order entities.

The evaluation of the genuine four-loop master integrals is a very challenging task that involves a mix of analytic and numerical methods, with only a handful of cases completed so far \cite{Aapeli}. An important complementary tool, which holds for the simultaneous evaluation of all IR convergent four-loop integals, is based on the loop-tree duality (LTD) method of vacuum perturbative QFT \cite{Capatti:2019ypt}. This method was recently generalized to nonzero chemical potentials and successfully applied to the evaluation of the infamous "bugblatter" diagram \cite{Navarrete:2024zgz},\footnote{For the etymology of the name of this diagram, see \cite{Blaizot:2001vr}.} which represents the LO difference between the pressures of cold and dense QCD and its phase-quenched version (see, e.g., \cite{Moore:2023glb}). The main difference between the LTD and cutting rules methods is that in the former, one first evaluates all temporal momentum integrals instead of just the fermionic ones, which avoids the generation of artificial IR divergences and at least in some cases leads to more manageable numerical calculations.

\textbf{Soft and mixed sectors:} The only soft scale present in cold and dense QM originates from the long-distance screening of gluon fields and can be identified with the $T\to 0$ limit of the chromoelectric screening mass $m_\text{E}\sim g\mu_B$ in eq.~(\ref{mE}), i.e.~the same parameter that appears in the LO HTL self energy derived in the previous section. At the $O(\alpha_s^3)$ order for the pressure, it no longer suffices to only work with this quantity, but we need to additionally consider other elements of the effective theory, including the HTL vertex function, and derive corrections to the LO self energy both in the form of two-loop contributions and power corrections in the soft external momentum. The necessary calculations were first described in \cite{Gorda:2021kme} and subsequently carried out in \cite{Gorda:2021znl,Gorda:2023mkk}, to which we refer the interested reader for technical details. Here, we instead restrict our attention to a qualitative account of the origins of the soft and mixed $O(\alpha_s^3)$ contributions to the QM pressure, see eq.~(\ref{presschem}).

\begin{figure}[t]
\centerline{%
\includegraphics[width=8.5cm]{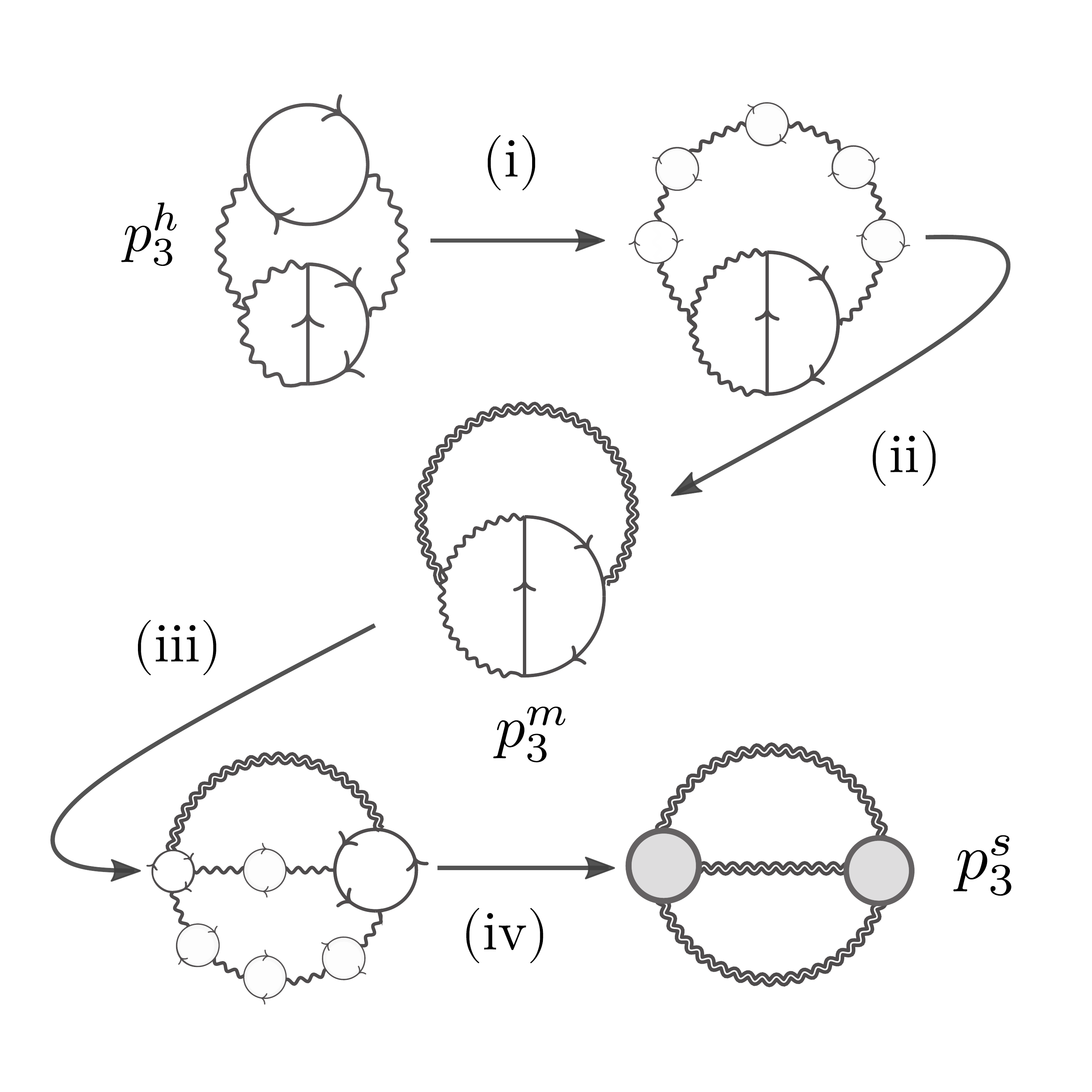}}
\caption{An illustration of how a single four-loop vacuum diagram in dense QCD gives rise to a three-loop mixed diagram and a two-loop HTL graph when one or several of the internal gluon lines becomes soft. The thin wavy lines correspond to unresummed and the thick wavy lines to HTL-resummed gluon propagators, respectively, while the grey blobs stand for HTL-resummed vertex functions.}
\label{scales}
\end{figure}

The physical nature of the mixed and soft contributions to the  $O(\alpha_s^3)$ pressure can be neeatly illustrated through a practical example depicted in fig.~\ref{scales}, borrowed from \cite{Gorda:2021znl}. Here, we begin from a single unresummed four-loop vacuum diagram of full QCD that consists of two quark loops and four additional gluon propagators. The fermion propagators are by construction "hard", being characterized by the scale $\mu$ present in their temporal momentum components, while the gluon momenta can be either hard, with $P\sim \mu$, or soft, with $P\sim m_\text{E}$. Should one or several of the gluonic momenta in the graph --- say the vertical ones flowing into the one-loop self-energy insertion in the first diagram of fig.~\ref{scales} --- become soft, a simple power-counting exercise shows that similar diagrams containing an arbitrary number of one-loop gluon self-energy insertions contribute at the same order in the weak-coupling expansion [step (i) in fig.~\ref{scales}]. Summing all such diagrams using the low-momentum limit of the gluon self energy [step (ii)] leads to the emergence of a three-loop graph containing an HTL resummed gluon propagator, which represents a typical mixed contribution to the $O(\alpha_s^3)$ pressure, denoted by $p_3^m$ in eq.~(\ref{presschem}). Repeating finally the same exercise for the two remaining unresummed gluon lines, which clearly become both soft at the same time  [step (iii)], leads to a fully soft two-loop contribution computable within the HTL effective theory  [step (iv)]. The final HTL graph  features also twp resummed HTL vertex functions, one of which arises from the lower quark loop of the original diagram, and contributes to the $p_3^s$ term of eq.~(\ref{presschem}).

The successful evaluation of the soft and mixed contributions to the $O(\alpha_s^3)$ pressure presents a technically very challenging task that was completed only recently in \cite{Gorda:2021znl,Gorda:2023mkk}. In these calculations, it was crucial to ensure the proper cancellation of all IR divergences in the hard four-loop diagrams of QCD against UV divergences originating from the soft two-loop diagrams in the HTL effective theory as well as UV and IR divergences originating from the mixed three-loop graphs. It is also worth repeating that, as noted above, the one-loop HTL self energies needed be generalized both by two-loop corrections, evaluated in \cite{Gorda:2023zwy}, and by power corrections in the soft external momentum, derived in \cite{Gorda:2023mkk}.

\begin{figure}[t]
\centerline{%
\includegraphics[width=10.5cm]{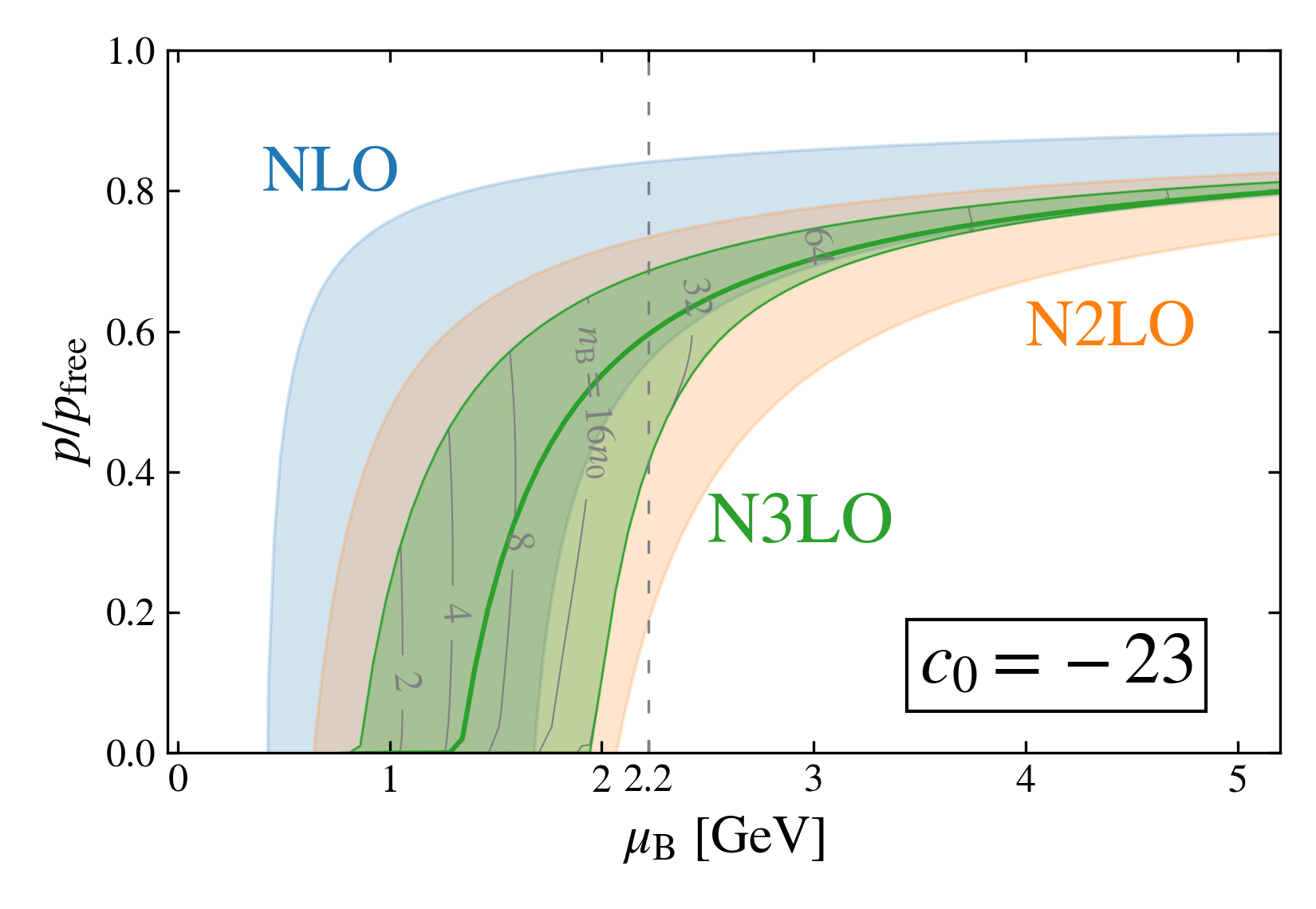}}
\caption{The pressure of cold and dense massless QM in beta equilibrium \cite{Gorda:2023mkk}. Shown are the NLO (blue band), NNLO (yellow), and NNNLO (green) perturbative orders, of which the last one is new and contains one single undetermined coefficient $c_0$ set to a value determined by a Bayesian estimation performed in \cite{Gorda:2023mkk} (see also \cite{Gorda:2023usm}).}
\label{pRisto}
\end{figure}

\textbf{Convergence of the weak-coupling expansion:} After the recent determination of the mixed $O(\alpha_s^3)$ contributions to the QM pressure in \cite{Gorda:2023mkk}, the current state-of-the-art result for the EoS of cold and dense QM contains all terms appearing in eq.~(\ref{presschem}) except for the finite, renormalization-scale-independent part of $p_3^h$, which amounts to one single number. As demonstrated in fig.~\ref{pRisto}, taken from \cite{Gorda:2023mkk}, the effect of this new order on the convergence of the weak-coupling expansion is dramatic. For typical values of the missing coefficient, currently under intense scrutiny, the new N3LO band improves the convergence of the pressure at all relevant densities. Such a result would importantly enable lowering the density, at which the high-density constraint is applied in model-independent determinations of the NS-matter EoS, from the present approx.~40$n_s$ down to perhaps $20n_s$. As we will discuss in detail in sec.~7 below, this would have a dramatic effect on the precision, at which we currently understand the properties of the ultradense matter present inside NS cores.

\section{Transport properties of dense QM}

In addition to the bulk thermodynamic properties of NS matter, first-principles particle theory methods may provide valuable input for its transport coefficients, describing physical phenomena such as diffusion, energy and momentum dissipation, and the conduction of heat and electric currents. While in some cases also relevant for the properties of quiescent NSs, they play a particularly important role in more dynamical settings, such as supernovae explosions and the postmerger dynamics of binary NS mergers (see, e.g., \cite{Radice:2021jtw,Chabanov:2023blf}). In \cite{Alford:2017rxf}, it is argued that the most important transport coefficient for NS mergers is the bulk viscosity $\zeta$, which has subsequently received considerable attention in recent literature, including two very recent papers \cite{CruzRojas:2024etx,Hernandez:2024rxi} concentrating on its determination in QM. Here, we will, however, begin the story with some  general remarks followed by a brief look into other transport coefficients in (mostly unpaired) dense QM, returning to the somewhat special evaluation of $\zeta$ only at the very end of this section.

Similarly to the more widely studied case of hot QGP, the transport coefficients of cold and dense QM are technically considerably more challenging to determine than its equilibrium thermodynamic properties, which is mainly due to the  inherently Minkowskian nature of the former quantities. This is reflected in the existence of only a handful of low-order perturbative results for the most important transport coefficients, many of which moreover date back tens of years \cite{Madsen:1992sx,Heiselberg:1993cr}. These quantities are typically defined through the response of the system to an imbalance in some local quantity, and their evaluation often begins from the Boltzmann equation, describing the evolution of an off-equilibrium statistical system. Here, our treatment of the subject will be rather superficial, and the interested reader is referred to the excellent review article \cite{Schmitt:2017efp} for more details on the transport pheanomena of dense QCD matter, including its confined phase. 

The study of  transport phenomena in dense QM often begins from the Boltzmann equation describing the behavior of the quark distribution function $f(\mathbf{r},\mathbf{p},t)$,
\begin{eqnarray}
\bigg(\frac{\partial}{\partial t}+\mathbf{v}\cdot\nabla_\mathbf{r}+\mathbf{F}\cdot\nabla_\mathbf{p}\bigg)f(\epsilon_\mathbf{p}) &=& \bigg(\frac{\partial f(\epsilon_\mathbf{p})}{\partial t}\bigg)_\text{coll}\, .  
\end{eqnarray}
Here, $\mathbf{F}$ is an external force, $\mathbf{r}$ and $\mathbf{p}$ denote the position and three-momen\-tum of the quark being tracked, and the  gluon-mediated scattering between quarks is described by the highly nonlinear collision term on the right-hand side of the equation (see, e.g., eq.~(1) of \cite{Heiselberg:1993cr}). The near-equilibrium solutions of this equation under various boundary conditions, say with a specific type of external force or a fixed flow velocity for one quark flavor, inform us about the response of the system to the perturbation in question and often allow solving for a particular transport coefficient.

A very crude approximation to solving the Boltzmann equation amounts to linearizing the collision term in the difference between the off-equilibrium and equilibrium distribution functions. This amounts to the so-called relaxation time approximation, which was applied in the first works on transport in dense QM or hot QGP, such as \cite{Hosoya:1983xm}. A major step forward was taken somewhat later in \cite{Heiselberg:1993cr}, where the first self-consistent perturbative study of transport in dense QM was performed with the full nonlinear collision term. In this work, a major challenge was related to the long-range interactions mediated by soft gluons, which  --- analogously to the $O(\alpha_s^2)$ pressure discussed above --- would lead to divergent results if  screening effects are not properly taken into account. 

While present also at high temperatures, the effects of dynamical screening, such as Landau damping, are particularly pronounced in transport calculations at large chemical potentials and small temperatures, where the system is characterized by three distinct momentum scales, $\mu_B$, $T$, and $m_\text{E}$. Of the three, the first can typically be assumed to be the  largest, but the ordering of the latter two is \textit{a priori} unclear in the physical systems of interest, such as binary NS mergers. This leads to technical complications in the Boltzmann description, where the quark-quark scattering matrix element within the collision term needs to be dressed with HTL-resummed gluon propagators, featuring the self-energies of eqs.~(\ref{HTLT}) and (\ref{HTLE}). 

From this point onwards, the steps needed to obtain LO results for various transport coefficients depend strongly on the quantity in question, and we refer the interested reader to \cite{Heiselberg:1993cr} for computational details. Here, we merely note that two key quantities that aid the determination of a number of transport coefficients from the diffusion constant to various conductivities and the shear viscosity include the momentum transfer between quarks and the so-called momentum stopping time. Since the publication of these LO results in 1993 \cite{Heiselberg:1993cr}, no perturbative improvements have, however, been reported, which is largely due to the technical complexity of the NLO calculations, reflected also in the first appearance of NLO transport results for high-temperature QGP as late as in 2018 \cite{Ghiglieri:2018dib,Ghiglieri:2018dgf}. Promising advances have, however, been recently made in the holographic description of the same transport coefficients as well as neutrino transport in the strongly-coupled regime of QCD-like theories \cite{Hoyos:2020hmq,Hoyos:2021njg,Jarvinen:2023xrx}.


Returning finally to the transport coefficient of highest relevance for NS mergers, the bulk viscosity $\zeta$ of dense QM, we encounter a quantity whose evaluation differs dramatically from the computations discussed above. Interestingly, it turns out that in a NS setting the dominant contribution to $\zeta$ does not originate from QCD alone, but emerges when rapid density oscillations in the macroscopic system force matter to depart from chemical equilibrium due to weak interactions not being able to keep up with the compression rate. In rotating quiescent pulsars, these oscillations are related to the different oscillation modes of the star (see, e.g., \cite{Kruger:2021zta}), while in binary NS mergers they are often related to the complicated merger dynamics. A straightforward calculation reviewed, e.g., in Appendix A of \cite{CruzRojas:2024etx} shows that in the so-called neutrino-transparent regime, where the dominant contribution to the bulk viscosity comes from the non-leptonic W-boson exchange process $u+d\leftrightarrow u+s$, the bulk viscosity obtains the schematic form 
\begin{eqnarray}
\zeta&=&\frac{\lambda_1 A_1^2}{\omega^2+(\lambda_1 C_1)^2}\, . \label{zeta}
\end{eqnarray}
Here, $\omega$ is the angular velocity of the density oscillations, $\lambda_1$ an electroweak rate defined by $\frac{\text{d}n_d}{\text{d}t}=-\frac{\text{d}n_s}{\text{d}t}=\lambda_1(\mu_s-\mu_d)$, and the coefficients $A_1$ and $C_1$ are linear combinations of quark densities and susceptibilities, reproduced in eqs.~(4) and (5) of \cite{CruzRojas:2024etx}. Interestingly,  $A_1$ turns out to identically vanish for mass-degenerate quarks, reflecting the vanishing of the bulk viscosity for conformal systems and implying that it is necessary to include a nonzero strange quark mass in the evaluation of the coefficients $A_1$ and $C_1$.

\begin{figure}[t]
\centerline{%
\includegraphics[width=11cm]{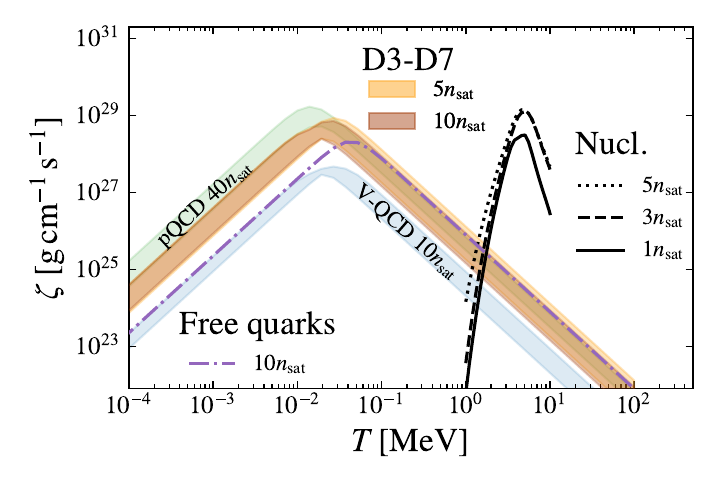}}
\caption{The $T$-dependence of the bulk viscosity in three-flavor unpaired QM  \cite{CruzRojas:2024etx}, evaluated in pQCD at the baryon density of $40n_s$ and in two holographic models V-QCD \cite{Jarvinen:2011qe,Jokela:2018ers} and D3-D7 \cite{Karch:2002sh} at 5 and $10n_s$. For reference, the quantity is also shown for dense nuclear matter at three different densities \cite{Alford:2022ufz} and at $n_B=10n_s$ for a system of noninteracting quarks.}
\label{bulk}
\end{figure}

Barring loop corrections to the electroweak data $\lambda_1$ (that may well be sizable, see e.g.~\cite{Schwenzer:2012ga}), a crucial implication of the form of eq.~(\ref{zeta}) is that the QCD contribution to the bulk viscosity of dense QM enters solely through the thermodynamic quantities appearing in the coefficients $A_1$ and $C_1$. As noted above, in their evaluation it is imperative to employ a nonzero strange quark mass, which may, however, be treated as an interaction following, e.g., the mass-expansion scheme recently introduced in \cite{Gorda:2021gha}. Precisely this was recently done in \cite{CruzRojas:2024etx}, where the thermodynamic functions contributing to the bulk viscosity of unpaired QM were evaluated to an unprecedented accuracy using both pQCD and two holographic models. The result of this computation is displayed in fig.~\ref{bulk}, where the bulk viscosity is plotted as a function of $T$ for various fixed baryon densities. We observe a qualitative agreement between the weak- and strong-coupling calculations within QM --- even across different densities --- but a sharp contrast to earlier nuclear-matter results from \cite{Alford:2022ufz}, for which the bulk viscosity peaks at a considerably higher temperature. It is tempting to speculate, whether this difference in the peak temperatures might lead to observable effects in the postmerger GW spectrum of a binary NS merger, should the effects of bulk viscous dissipation on the merger dynamics be sizable enough.

One general remark about transport coefficients in dense QM is finally in order. As the alert reader has surely noted, all of the discussion above concerned the transport properties of unpaired QM, which is a natural starting point, but very likely not the physical ground state of QCD at the low temperatures and moderate and high densities realized in NSs and their binary mergers. Instead, it is widely believed that some type of quark pairing takes place in physical QM, but due to the strongly coupled nature of the system the details of the pairing channel remain unclear at non-asymptotic densities. Unlike in the case of equilibrium thermodynamics, where pairing contributions to the EoS are expected to be strongly suppressed at all densities where pQCD is applicable, many transport coefficients are moreover thought to be highly sensitive to pairing, so calculations performed in the unpaired phase will need to be generalized to color-superconducting phases in the future. Given the technical nature of this topic, we refrain from a more extensive discussion here and instead refer the interested reader to two excellent review articles by Andreas Schmitt and collaborators \cite{Alford:2007xm,Schmitt:2017efp}.

\section{Model-independent inference of NS-matter properties}

After describing the determination of the thermodynamic and transport properties of unpaired QM at some length above, it is natural to finally ask, how these results can be used in a more fenomenological setting within NS physics. Here, the primary interest lies in a model-independent constraining of the EoS and other properties of strongly interacting matter up to the maximal $O(5-10n_s)$ densities reached inside physical NSs and the maximal $O(50-100\text{MeV})$ temperatures reached during NS mergers. \textit{A priori}, it is far from clear that perturbative results of the kind described in these notes contain any information of practical use in such settings, given that weak-coupling expansions in pQCD typically start converging only at densities many times higher. Somewhat counterintuitively, the perturbative high-density constraint, however, turns out very useful especially in EoS inference, and NS observations can in turn be shown to constrain the thermodynamic properties of QCD matter up to relatively high densities. Below, we will shed light on these developments, paying particular attention on the use of the pQCD constraint in the model-independent inference of NS-matter properties. For a reader interested in NS physics from a more astrophysical point of view, we can recommend, e.g., the recent review \cite{Nattila:2022evn}.

The key microscopic quantity one typically tracks in the description of quiescent NSs is the EoS of dense QCD matter in the limits of vanishingly small temperature $T$, local charge neutrality,\footnote{The requirement of local charge neutrality can in principle be relaxed to a global one, allowing for the presence of two phases in coexistence, relevant for the case of a first-order (deconfinement) phase transition. This is referred to as the Gibbs construction, explained in more detail in, e.g., \cite{Han:2019bub}.} and beta equilibrium. This is because it is precisely the relationship between the pressure and energy density of this type of matter that closes the famous Tolman-Oppenheimer-Volkov (TOV) equations governing hydrostatic equilibrium inside a nonrotating star \cite{Tolman:1939jz,Oppenheimer:1939ne}, 
\begin{eqnarray}
\frac{dM(r)}{dr} &=& 4\pi r^2 \epsilon(r)\, , \nonumber \\
 \frac{dp(r)}{dr}  &=& -\frac{(\epsilon(r) + p(r))}{r^2}  \frac{ (M(r) + 4\pi r^3 p(r)  )}{  ( 1-2M(r) /r )}\, ,
\end{eqnarray}
allowing one to solve the corresponding unique mass-radius (MR) relation that can be compared to observations. Up to small deviations caused by rotational frequencies, magnetic fields, and other similar quantities varying from star to star, the masses and radii of every NS in existence should in principle fall on the same MR-curve, dictated by the low-temperature EoS alone. Conversely, should an accurate simultaneous determination of the masses and radii of several individual NSs become feasible one day, one may reverse engineer the ToV equations to obtain the EoS of cold and dense beta-equilibrated QCD matter all the way from vanishing density to the maximal central densities realized in stable NSs.

\begin{figure}[t]
\centerline{%
\includegraphics[width=13.0cm]{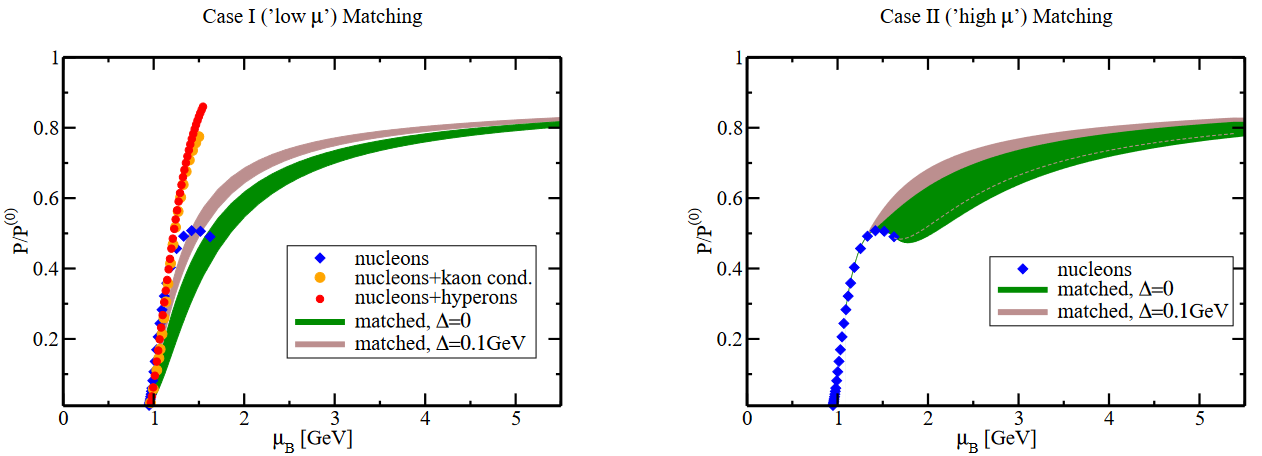}}
\caption{Results from a naive matching of phenomenological hadronic EoSs (blue, red, and yellow dots and diamonds) to the perturbative QM EoS both with (brown band) and without (green) a phenomenological pairing contribution \cite{Kurkela:2009gj}.}
\label{coldQM}
\end{figure}

The most straightforward way, in which one could in principle use the perturbative QM EoS in NS physics, is by extrapolating the result to much lower chemical potentials, where it would be matched to a nuclear matter EoS,  similarly extrapolated from below. The simplest implementation of such a setup would typically involve a first-order phase transition at the baryon chemical potential where the two pressures are equal, while a more refined setting would allow for the presence of a mixed phase of both quark and nuclear matter. Precisely this was done in \cite{Kurkela:2009gj,Kurkela:2010yk}, where the three-loop pressure of cold QM was first evaluated with a nonzero strange-quark mass and then matched with phenomenological EoSs for high-density NM (see fig.~\ref{coldQM}). This represented a significant step forward from earlier calculations employing simplistic model EoSs for QM, such as that of the MIT bag model \cite{Alcock:1986hz}, but the approach nevertheless suffered from multiple problematic issues. First, the 3-5$n_s$ densities, where the matching of the NM and QM EoSs was performed, are well outside the realms of controlled first-principles calculations in both phases, so that predictions for quantities such as the transition density or the associated latent heat can be considered indicative at best. Also, there is no robust way to quantitatively assess the systematic uncertainty involved in such a setup, and potentially sizable contributions from important physical phenomena such as the onset of hyperons \cite{Weissenborn:2011ut} or quark pairing \cite{Alford:2007xm} are altogether ignored.

The most important shortcoming of the above setup clearly originates from the assumption that the low- and high-density EoSs and their respective uncertainty ranges are applicable also in the problematic region of NS core densities. To improve from here, a natural alternative is to divide the density interval from zero to infinity to not two but three parts: a low-density regime where \textit{ab-initio} nuclear-physics methods such as CET produce reliable results, a high-density one where the pQCD EoS for QM converges in a satisfactory manner, and an intermediate region where no first-principles results are available. In the last density interval, the EoS can be allowed to behave in any physically consistent way, with only subluminality ($c_s<1$), thermodynamic consistency as well as a smooth matching to the low- and high-density EoSs required. 

\begin{figure}[t]
\centerline{%
\includegraphics[width=6.5cm]{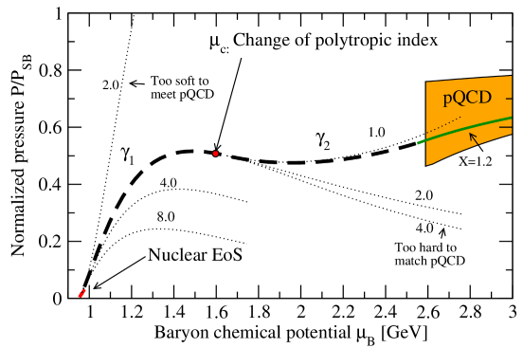}$\;\;\;$\includegraphics[width=6.2cm]{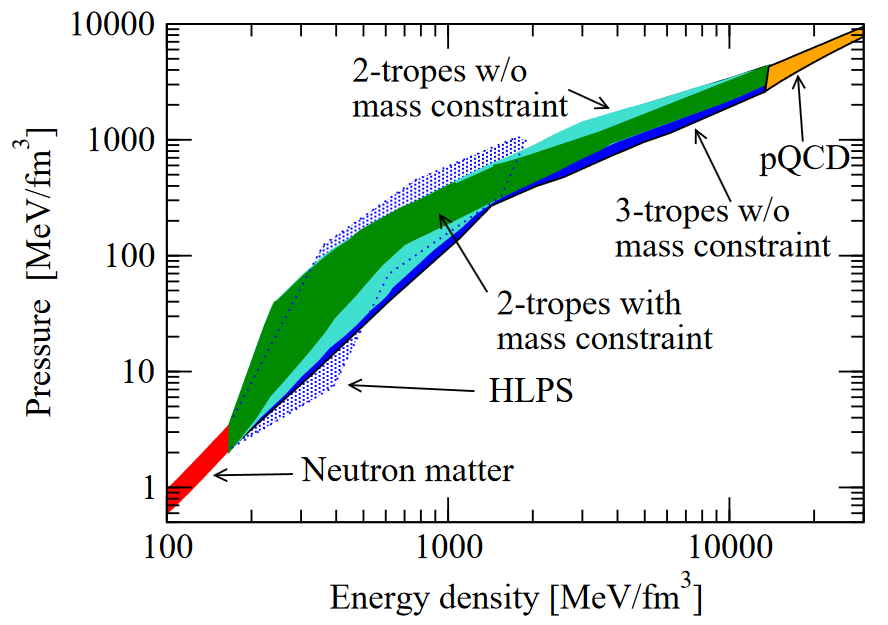}}
\caption{Left: an illustration of the schematic idea behind the first model-independent interpolation study of the NS-matter EoS \cite{Kurkela:2014vha}. The red line on the left illustrates the low-density EoS from CET \cite{Hebeler:2013nza}, the orange band on the right the high-density EoS from pQCD \cite{Kurkela:2009gj,Fraga:2013qra}, and the dashed black line in between represents a two-part polytrope. Right: the resulting EoS bands from the study presented in \cite{Kurkela:2014vha}. Shown here are bands generated with two- (green) and three-part (solid blue) polytropes as well as results from an earlier extrapolation study \cite{Hebeler:2013nza} dotted blue) and a two-part polytrope calculation without a constraint requiring that two-solar-mass NSs be supported by all viable EoSs (turquoise).}
\label{fksv}
\end{figure}

The qualitative idea described above and illustrated in fig.~\ref{fksv} was first implemented by Kurkela et al.~in 2014 \cite{Kurkela:2014vha}, following a similar extrapolation of the CET EoS by Hebeler et al.~just one year before \cite{Hebeler:2013nza}. The exploratory study of \cite{Kurkela:2014vha} involved the use of two- and three-part polytropes in the intermediate density regime, i.e.~functions composed in a piecewise manner from ans\"atze of the form $p=\kappa n_B^\gamma$. After discarding EoS specimens that do not support the existence of the most massive NSs observed, $M\approx 2M_\odot$ with $M_\odot$ standing for the solar mass, the resulting EoS band was seen to considerably tighten from that obtained in \cite{Hebeler:2013nza} (see fig.~\ref{fksv}). In addition, a set of somewhat \textit{ad-hoc} limits imposed on the polytropic $\gamma$ indices in the extrapolated EoSs of \cite{Hebeler:2013nza} were now set on more firm footing and shown to originate from the high-density pQCD constraint.

Without significant advances in the astrophysical measurements that can be applied to constrain the interpolated EoS band at intermediate densities, tightening the EoS bounds derived in \cite{Kurkela:2014vha} would have been a very slow process, driven solely by advances in the low- and high-density microphysical calculations. In late 2017, a dramatic new observation was, however, announced by the LIGO and Virgo collaborations, who reported the first-ever detection of a GW signal from the binary NS merger GW170817 \cite{LIGOScientific:2017vwq,LIGOScientific:2018cki,LIGOScientific:2018hze}. The GW signal was soon accompanied by associated gamma-ray and kilonova obervations \cite{LIGOScientific:2017ync,LIGOScientific:2017zic}, suggesting that the merger remnant very likely underwent a delayed gravitational collapse into a black hole (BH).

\begin{figure}[t]
\centerline{%
\includegraphics[width=\textwidth]{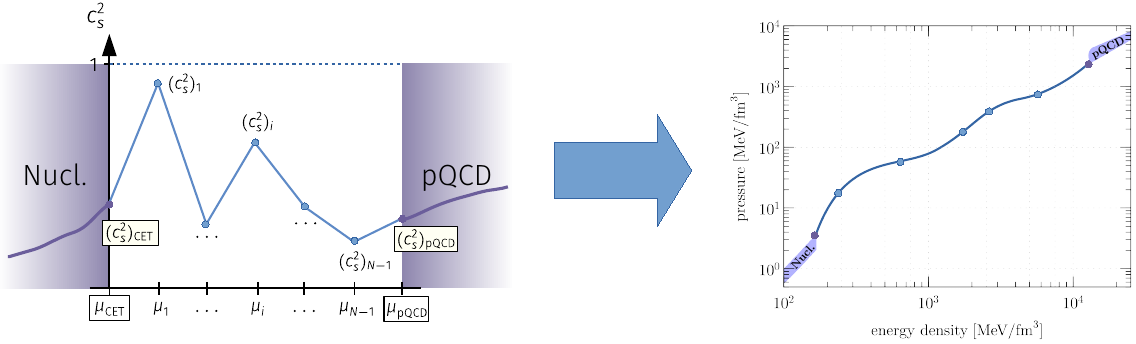}}
\caption{A schematic illustration of the speed-of-sound interpolation routine, originally developed for \cite{Annala:2019puf}. The figure is taken from \cite{Annala:2021gom}.}
\label{cartoon}
\end{figure}

From the point of view of EoS inference, the most useful constraint from the first LIGO/Virgo papers was a nontrivial upper bound for the tidal deformability of an approx.~$1.4$-solar-mass NS --- a quantity that characterizes the extent, to which the shapes of two inspiraling stars get deformed under each others' gravitational fields \cite{LIGOScientific:2018cki}. With accurate knowledge of the so-called chirp mass of the merger, this bound was quickly translated to a constraint for the NS-matter EoS, whereby a large fraction of otherwise viable candidate EoSs could be discarded for predicting too large radii and thus tidal deformabilities for light pulsars. Early papers implementing this new bound in either model-independent EoS studies or specific model EoSs included e.g.~\cite{Annala:2017llu,Margalit:2017dij,Rezzolla:2017aly,Ruiz:2017due,Bauswein:2017vtn,Radice:2017lry,Most:2018hfd,Dietrich:2020efo,Capano:2019eae,Landry:2018prl,Raithel:2018ncd,Raithel:2019ejc,Raaijmakers:2019dks,Essick:2019ldf,Jokela:2020piw,Al-Mamun:2020vzu,Demircik:2020jkc}, of which the one by Annala et al.~\cite{Annala:2017llu} was the first one to systematically utilize the pQCD high-density constraint.

\begin{figure}[t]
\centerline{%
\includegraphics[width=5.6cm]{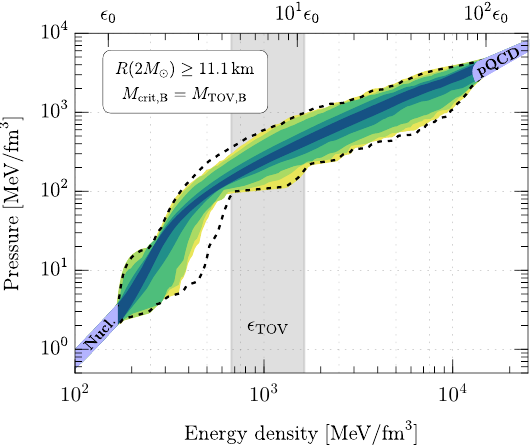}$\;\;\;\;$\includegraphics[width=6.8cm]{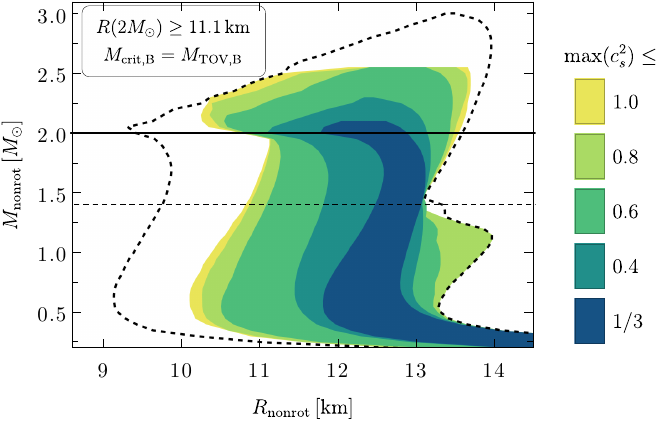}}
\caption{An example of the EoS and MR results from the analysis of \cite{Annala:2021gom}, where information from the NS radius constraints from the NICER collaboration and the likely delayed gravitational collapse of the GW170817 merger remnant were taken into account, assuming here the more conservative supramassive scenario. The color coding refers to the maximal value the speed of sound squared attains at any density, made feasible by the $c_s^2$ interpolation routine of fig.~\ref{cartoon}. The observed behavior clearly suggests that a hypothetical upper limit for $c_s^2$ (see, e.g., the recent \cite{Hippert:2024hum}) would carry significant additional constraining power for EoS inference.}
\label{LIGO1}
\end{figure}

Another important insight that emerged from the GW170817 merger was related to the likely formation of a black hole through either a so-called supramassive or hypermassive NS (see, e.g., \cite{Baiotti:2016qnr} for an extensive discussion of these scenarios).\footnote{In short, in the hypermassive NS scenario, the remnant experiences a rapid (milliseconds to a few seconds) gravitational collapse while still undergoing differential rotation, while in the longer-lived supramassive NS scenario, the remnant has time to turn from differential to uniform rotation, then slowly loses angular momentum, and finally collapses to a BH in a timeframe from seconds to hours.} Depending on the scenario considered, this leads to either more stringent (the more likely hypermassive scenario) or weaker (supramassive case) constraints that were implemented in an interpolation setup featuring the pQCD EoS in \cite{Annala:2021gom} (see also \cite{Margalit:2017dij,Rezzolla:2017aly,Ruiz:2017due}, which derived the first constraints on the maximal TOV mass of a nonorotating NS and \cite{Fan:2023spm} for a later refinement). The corresponding interpolation routine started from expressing the speed of sound squared as a piecewise-linear function of the quark chemical potential, illustrated in fig.~\ref{cartoon}, and resulted in the EoS and MR bands of fig.~\ref{LIGO1}. In the latter results, radius constraints published by the NICER collaboration in 2021 were implemented as well, amounting to an 11.1km lower bound for the radius of a $2M_\odot$ NS at a 90\% credence \cite{Miller:2021qha,Riley:2019yda,Riley:2021pdl}.

While the results of fig.~\ref{LIGO1} still represent the state-of-the-art of hard-cut-type interpolation analyses, there are questions that studies of this kind are unable to answer or that would at least require a further layer of analysis: 
\begin{enumerate}
\item What is the quantitative impact of the high-density constraint on the results, i.e.~would similar results have been obtained also without the pQCD EoS that we have worked so hard to derive in these notes? 
\item Is there a way to assign likelihoods for various EoS behaviors and take measurement uncertainties into account? The use of hard cuts in EoS inference is clearly suboptimal, as it forces one to make a binary decision (accept or reject a given EoS) based on minuscule differences, such as whether the most massive NS an EoS is able to support is $1.99M_\odot$ or $2.00M_\odot$.
\item Is it possible to transform the EoS constraints to statements about the  phase of matter inside NS cores? In particular, can we somehow estimate the likelihood of NS cores containing deconfined QM? 
\end{enumerate}
Below, we will address each of these questions in turn, covering both the technical tools needed in this work and the results from the said analyses.

Starting from the first point, one may in principle ask two slightly different questions: 1) Do the piecewise-defined interpolation functions used in calculations such as \cite{Annala:2017llu,Annala:2019puf,Annala:2021gom} allow for sufficient complexity for the EoS, so that the results are not biased by the EoS ans\"atze? and 2) How would the results change if the corresponding analyses were performed without the high-density constraint? To address both issues at one go, Oleg Komoltsev, Aleksi Kurkela and collaborators decided to inspect the impact of the pQCD constraint with an entirely different set of tools. In \cite{Komoltsev:2021jzg}, they first demonstrated that the high-density constraint implemented at $40n_s$ limits the NS-matter EoS down to below $3n_s$ with no assumptions except for causality and standard thermodynamic relations. Following this, they employed a very conservative nonparametric Gaussian Process (GP) interpolation to pinpoint the effefts of the pQCD constraint on the NS-matter EoS \cite{Gorda:2022jvk,Komoltsev:2023zor}. The most important takeaway from these studies, where the pQCD limit was treated in a similar fashion to astrophysical constraints and could be turned on and off at will, was that it significantly softens the EoS at intermediate and high densities, leading to a lower maximum mass for stable NSs and a higher likelihood for the formation of a BH in binary NS mergers. In \cite{Komoltsev:2023zor}, the authors in addition introduced a less conservative method for implementing the pQCD constraint to EoS inference calculations, with the corresponding marginalized QCD likelihood function  available at \url{https://zenodo.org/records/10592568}.

As to the second question concerning likelihoods and measurement uncertainties, a very natural step forward is to extend the hard-cut analyses described above towards Bayesian statistical inference, built on the famous Bayes' theorem
\begin{eqnarray}
P({\rm EoS|data})&=&\frac{P({\rm data|EoS})P({\rm EoS})}{P({\rm data})}\, .
\end{eqnarray}
In short, the idea is to turn the (difficult) question of determining the relative likelihoods of various EoSs based on given observational data around and instead solve the (substantially easier) problem of determining the likelihood of the observational data based on a given EoS. Such a setup has by now been implemented to the NS-matter EoS inference by multiple groups, with early adaptations including, e.g., \cite{Raithel:2017ity,Coughlin:2018fis,Guven:2020dok,Malik:2022zol,Jiang:2022tps}. A key finding from these studies is the very likely presence of a bump, i.e.~a maximum, in the speed of sound of NS matter at densities realized within physical NSs.

Finally, succesfully answering the third and last of the above questions requires knowledge of the expected properties of both nuclear and quark matter in the strongly-coupled regime, where the deconfinement transition is expected to occur but no controlled first-principles tools are available. The unavailability of \textit{ab-initio} results makes the question somewhat ill-posed, but at least some guidance can be sought from a more tractable regime of high-energy-density QCD matter, high-temperature QGP. For this system, lattice simulations carried out by multiple collaborations over the past two decades (see, e.g., \cite{Borsanyi:2010cj,HotQCD:2014kol}) have convincingly demonstrated that the transition from a hot hadron gas to QGP takes place as a crossover transition at a temperature of approx.~155 MeV and an energy density of slightly below 400 MeV/fm$^3$, although the precise numbers slightly depend on the quantities being tracked. The most striking difference between the confined and deconfined phases is related to the approximative conformal symmetry of the latter: while the properties of hadronic matter are characterized by the $O($GeV$)$ hadron masses, in QM the only scaleful parameters are the $O($100 MeV$)$ mass of the strange quark and the dynamically generated $\Lambda_\text{QCD}\sim$ 300 MeV scale parameter. This is reflected in the rapid conformalization of lattice results for multiple quantities, when the temperature of the system is increased past the transition region, cf.~e.g.~figs.~7 and 11 of \cite{HotQCD:2014kol}.

For the cold and dense QCD matter inside NSs, the lack of first-principles predictions to compare against means that phase identification must at least partially rely on a comparison of the inferred properties of NS matter with the known properties of QM and NM at considerably higher and lower densities, respectively. Fortunately, there exist a large number of physical quantities with clear predictions in various limits, to which we can compare the results of our model-independent EoS inference. Such quantities include, e.g., the speed of sound squared $c_s^2$, the normalized trace anomaly and its logarithmic derivative with respect to energy density, $\Delta = (\epsilon-3p)/(3\epsilon)$ and $\Delta'\equiv {\rm d}\Delta/{\rm d}\ln \epsilon$, the "conformal distance" $d_c\equiv \sqrt{\Delta^2+(\Delta')^2}$, the polytropic index $\gamma= d\ln p/d\ln \epsilon$, as well as the normalized pressure $p/p_\text{free}$. In table \ref{table1}, taken from \cite{Annala:2023cwx}, we summarize the values these quantities take in various limits: sub-saturation-density CET calculations, model results for NM at NS-core densities, pQCD calculations around $n_B=40n_s$, field theories exhibiting exact conformal symmetry (CFTs), and systems undergoing a discontinuous first-order phase transition (FOPTs). A key takeaway from this table is that, as expected, high-density QM is considerably closer to the conformal limit than low- or high-density NM, implying that a possible conformalization of NS matter would be a strong indication of the presence of deconfined matter within NS cores.

\begin{table}[t]
\begin{tabular}{@{}l@{\hspace{4\tabcolsep}}r@{\hspace{4\tabcolsep}}r@{\hspace{4\tabcolsep}}r@{\hspace{4\tabcolsep}}r@{\hspace{4\tabcolsep}}r@{}}
\toprule
	& CET & Dense NM & Pert.~QM & CFTs & FOPT\\
\midrule
$c_s^2$	& $\ll 1$ & $[0.25,0.6]$ &$\lesssim 1/3$	&1/3  &0		\\
$\Delta$ & $\approx 1/3$	& $[0.05,0.25]$ &$[0,0.15]$	&0  & $1/3-p_\mathrm{PT}/\epsilon$		\\
$\Delta'$ & $\approx 0$	& $[-0.4,-0.1]$ &$[-0.15,0]$	&0  & $1/3-\Delta$		\\
$d_c$ & $\approx 1/3$	& $[0.25,0.4]$ &$\lesssim 0.2$	&0  & $\geq 1/(3\sqrt{2})$		\\
$\gamma$ &$\approx 2.5$	& $[1.95,3.0]$ &$[1,1.7]$	&1  &0		\\
$p/p_\text{free}$	&$\ll 1$ & $[0.25,0.35]$ & $[0.5,1]$	& --- & $p_\mathrm{PT}/p_\text{free}$		\\
\bottomrule
\end{tabular}

\caption{A summary of the values that several physical quantities take in various limits defined in the main text. The table is taken from \cite{Annala:2023cwx}.}
\label{table1} 
\end{table}

\begin{figure}[t]
\centerline{%
\includegraphics[width=12.0cm]{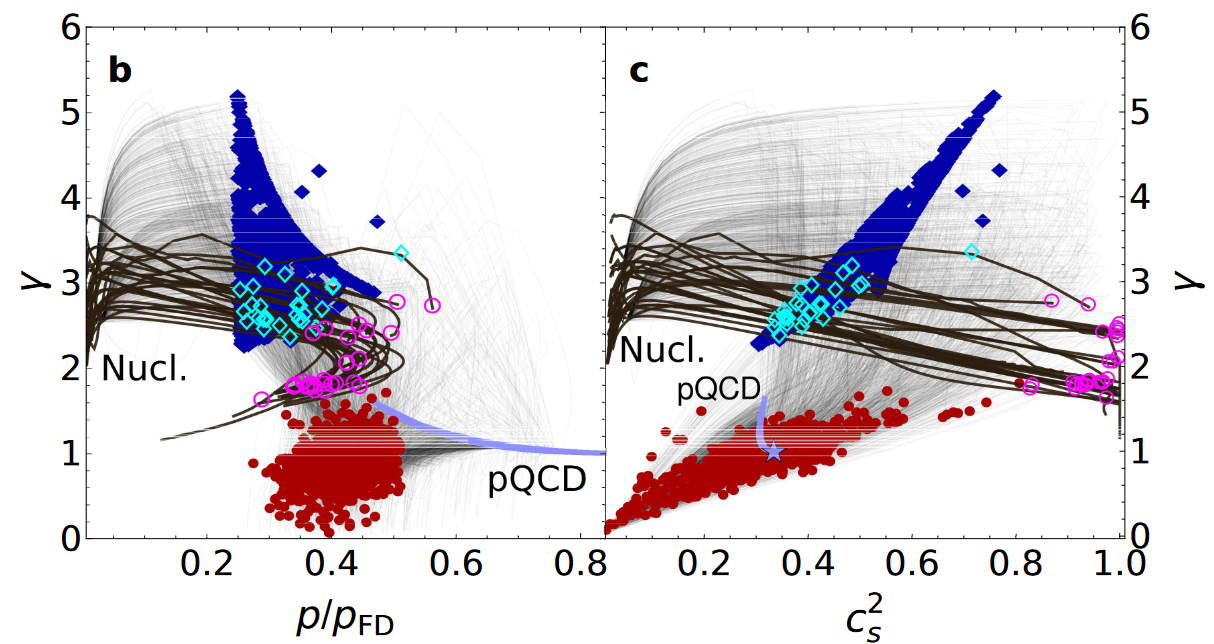}}
\caption{The model-independent NS-matter EoSs from \cite{Annala:2019puf} (thin grey lines) plotted as functions of normalized pressure, polytropic index, and speed of sound squared, and compared to the high-density pQCD limit (blue band) and viable models of high-density NM (thick black lines). Shown are also blue and cyan diamonds corresponding to the centers of $1.4M_\odot$ NSs according to the interpolated and model EoSs, respectively, and the same for maximally massive NSs (red and magenta circles).}
\label{QMcore1}
\end{figure}

The first serious attempt to inspect the possible conformalization of matter inside NS cores in a model-independent fashion was made in \cite{Annala:2019puf}, where the primary quantities studied included the polytropic index $\gamma$, the speed of sound squared $c_s^2$, and the normalized pressure $p/p_\text{free}$. The results that emerged from this hard-cut-type study are summarized in fig.~\ref{QMcore1}. In short, they indicate that while the properties of matter in the inner cores of light $1.4M_\odot$ NSs (blue and cyan diamonds in the figure) are well in line with those expected based on model calculations for dense NM, things look dramatically different in the cores of maximally massive NSs (red and magenta circles). In the latter case, the inferred properties of NS matter lie considerably closer to those of QM at perturbative densities, indicating the likely presence of deconfined matter inside the most massive stable NSs.

In the recent \cite{Annala:2023cwx}, the argument for QM cores was put on a considerably firmer footing by tracking the conformalization of a number of new physical quantities, utilizing the largest set of NS observations to date, and most importantly taking advantage of Bayesian inference techniques in the analysis. As an optimal quantity for tracking the conformalization of NS matter, the authors of \cite{Annala:2023cwx} chose the conformal distance $d_c\equiv \sqrt{\Delta^2+(\Delta')^2}$, which extends the use of the normalized conformal anomaly, first proposed in the NS context by Fujimoto et al.~\cite{Fujimoto:2022ohj}. The reasoning used was simple: while $\Delta$ may take on small values due to the quantity changing its sign around a given density, $d_c$ tends to zero only when both $\Delta$ and its rate of change are very small. In addition, when the triplet $\Delta$, $\Delta'$, and $d_c$ all tend to zero, then the closely related quantities $c_s^2$ and $\gamma$ will also approach their conformal values of $1/3$ and $1$ due to the relations
\begin{eqnarray}
\Delta &=& \frac{1}{3} - \frac{c_\mathrm{s}^2}{\gamma}\, , \quad
\Delta'  \;=\; c_\mathrm{s}^2 \biggl( \frac{1}{\gamma} - 1 \biggr)\, .
\end{eqnarray}

\begin{figure}[t]
\centerline{%
\includegraphics[width=6.1cm]{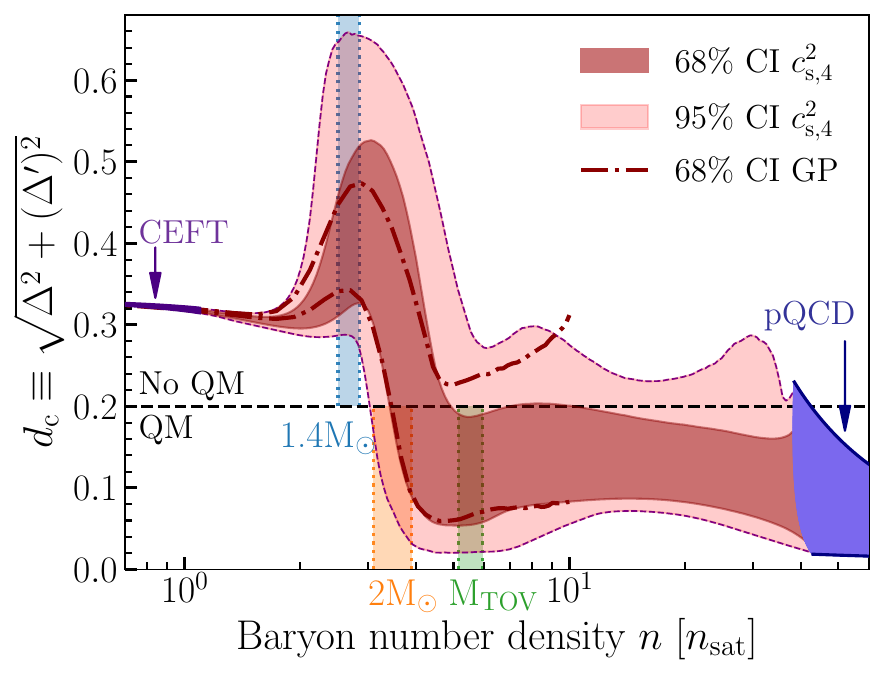}$\;\;$\includegraphics[width=6.5cm]{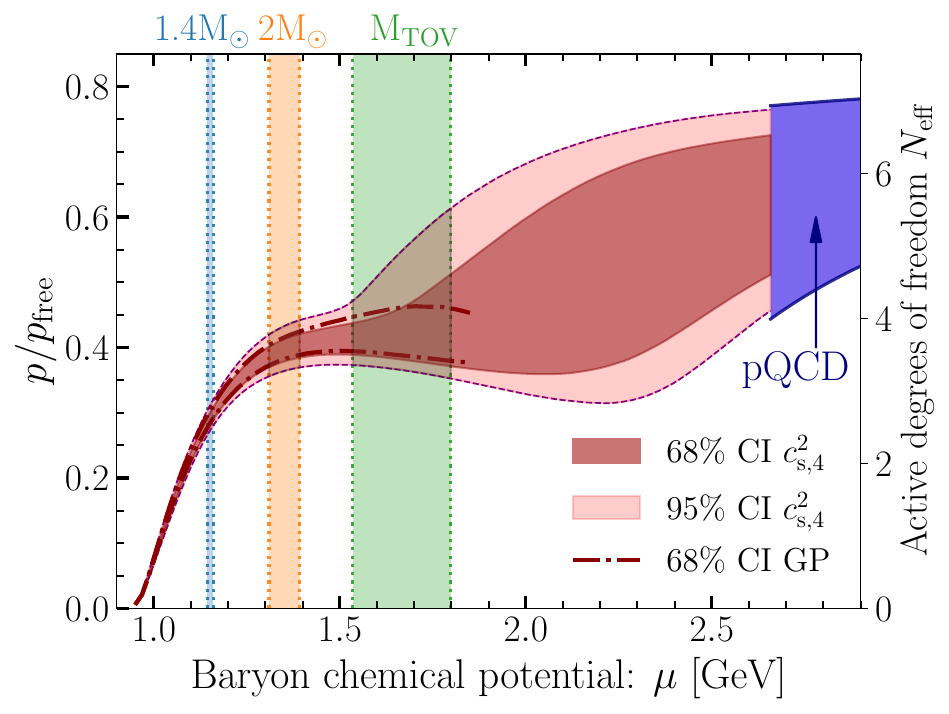}}
\caption{The behavior of the conformal distance $d_c$ and the normalized pressure $p/p_\text{free}$ in a recent Bayesian study of the NS-matter EoS \cite{Annala:2023cwx}. Shown are 68 and 95\% credible intervals from a speed-of-sound interpolation and the 68\% credible interval from a GP calculation, while the vertical bars indicate the central densities of $1.4M_\odot$, $2M_\odot$ and maximal-mass ($M_\text{TOV}$) NSs at the 68\% level.}
\label{QMcore}
\end{figure}

In fig.~\ref{QMcore} (left) from \cite{Annala:2023cwx}, we indeed witness the rapid conformalization of $d_c$ between the central densities of $1.4M_\odot$ and maximal-mass NSs. The corresponding likelihood for the existence of conformalized matter in the cores of the most massive stable NSs was found to be around 80-90\%, with the remaining 10-20\% likelihood corresponding to EoSs exhibiting FOPT-like behavior at the central densities of maximal-mass NSs, indicating the likely presence of a destabilizing phase transition  (see also \cite{Han:2022rug} for a related study with similar conclusions). Whether this observation can be interpreted as an indication for the existence of QM cores constitutes in principle a separate question, but some further reaffirmation can be obtained from the normalized pressure, shown in fig.~\ref{QMcore} (right), obtaining a value indicative of a QM-like number of effective degrees of freedom at the same densities. It is indeed worth noting that at high temperatures, the deconfinement transition has been shown to occur when $p/p_\text{free}\lesssim 0.2$ \cite{Borsanyi:2010cj,HotQCD:2014kol}, while in the cold and dense case the same quantity obtains values around 0.4 in the inner cores of maximal-mass stars. 

As noted already in \cite{Annala:2019puf,Annala:2023cwx} and alluded to above, one viable scenario still remains that would altogether forbid the existence of QM cores: a strong FOPT destabilizing the NS mass-radius sequence as soon as the central density of the star reaches the transition density. While it is nontrivial to make likelihood comparisons between two disjoint EoS ensembles --- e.g., the non-QM EoS in studies like \cite{Annala:2023cwx} corresponded to arbitrarily rapid crossovers --- this issue has been studied in recent literature to some extent. In the hard-cut study \cite{Gorda:2022lsk}, it was demonstrated that if a strong first-order transition is inserted in a simple polytropic NS-matter EoS, it is possible to construct EoSs that extend beyond the limits derived in papers such as \cite{Annala:2019puf,Annala:2021gom}. More recently, Oleg Komoltsev implemented FOPTs to a Bayesian framework involving GP regression \cite{Komoltsev:2024lcr}, finding no evidence either favoring or disfavoring destabilizing solutions. Other features seen in previous EoS inferences without explicit phase transitions were seen to stay intact, though, including most importantly a clear peak in the speed of sound. More work is clearly needed to confirm or rule out the presence of a destabilizing FOPT, possibly using a future postmerger GW signal that may carry traces of the presence of a discontinuous transition \cite{Most:2018eaw,Ecker:2024kzs}.

\section{Outlook to future developments}

In the lecture notes at hand, we have reviewed in some detail both the methods used in first-principles perturbative thermal field theory calculations within high-density quark matter, and the applications of these results to the model-independent determination of the neutron-star-matter equation of state. In this discussion, we have not aimed at a self-consistent text-book-style presentation, but have rather tried to aid the interested reader in their journey of self study. We have done so by filling in a number of gaps left uncovered in existing textbooks on thermal field theory and by providing numerous references to original research articles, including both older classics of the field and more recent studies.

The topic of these lecture notes belongs to a growing subfield of nuclear and particle physics, aimed at understanding the properties of strongly interacting matter inside NSs. This is a rapidly evolving research topic, not least due to the pace, at which neutron-star observations have progressed during the past 10-15 years. Many recent advances have been made possible by an efficient interplay between microscopic theoretical calculations and new observational insights, of which the quest to discover a new phase of QCD matter inside NS cores represents a prime example. Without \textit{ab-initio} limits for the properties of low-density nuclear matter and high-density QM, NS observations would at best provide ballpark estimates for the properties of NS interiors, and without new observational constraints, the accuracy of EoS inference would not have progressed much during the past decade.


Within the next couple of years, several important advances relevant for the physics of NSs and their potential QM cores can be expected to emerge from the microscopic side. CET and pQCD calculations of the thermodynamics of low-temperature QCD matter are advancing at a rapid pace, with the completion of the $O(\alpha_s^3)$ pressure of cold and dense QM being finally in sight. Once complete, this result is expected to dramatically improve the precision, to which we know the properties of deconfined matter at densities of order 15-40$n_s$. This will immediately have a significant impact on the model-independent inference of the NS-matter EoS, which continues to be an active topic of research \cite{Semposki:2024vnp,Malik:2024qjw,Tang:2024jvs}. Alongside such  developments, the transport properties dense QM are presently under intense scrutiny, and model-independent bounds for the temperature dependence of the NS-matter EoS, improving previous estimates such as those of \cite{Chesler:2019osn,Mroczek:2024sfp}, are upcoming. As reviewed in these lecture notes, a combination of such theoretical results and future observational advances, including the potential detection of a postmerger GW signal from a binary NS merger, are expected to hold the key to resolving long-standing puzzles in nuclear astrophysics.

\section*{Acknowledgments}

I would like to thank Niko Jokela, Aapeli K\"arkk\"ainen, Mika Nurmela, Risto Paatelainen, and Tomi Ruosteoja for useful comments on early versions of these lecture notes.




\printbibliography


\end{document}